\documentclass[aps,
pra,
twocolumn,
superscriptaddress,
10pt,
nofootinbib,
floatfix,
longbibliography]{revtex4-2}

\usepackage{multirow,amsthm,amssymb,amsbsy,amsmath,bm}

\usepackage{enumitem}
\usepackage{graphicx}
\usepackage{booktabs}
\usepackage{verbatim}
\usepackage{dcolumn}
\usepackage{color}
\usepackage[dvipsnames]{xcolor}

\usepackage{mathtools}
\usepackage[normalem]{ulem}
\usepackage{soul}
\usepackage[colorlinks=true,linkcolor=blue,citecolor=blue,urlcolor=blue]{hyperref}
\usepackage{comment}

\usepackage{amsmath}

\DeclareMathOperator*{\argmin}{arg\,min}

\usepackage{physics}

\newcommand{\mar}[1]{{\color{black}#1}}

\usepackage{orcidlink}

\begin{document}

\title{Tensor network noise characterization for near-term quantum computers}

\author{Stefano Mangini\,\orcidlink{0000-0002-0056-0660}}
\email{stefano.mangini@algorithmiq.fi}
\affiliation{Algorithmiq Ltd, Kanavakatu 3C 00160 Helsinki, Finland}
\affiliation{QTF Centre of Excellence, Department of Physics, University of Helsinki, P.O. Box 43, FI-00014 Helsinki, Finland.}

\author{Marco Cattaneo\,\orcidlink{0000-0001-6614-5286}}
\affiliation{Algorithmiq Ltd, Kanavakatu 3C 00160 Helsinki, Finland}
\affiliation{QTF Centre of Excellence, Department of Physics, University of Helsinki, P.O. Box 43, FI-00014 Helsinki, Finland.}

\author{Daniel Cavalcanti\,\orcidlink{0000-0002-2704-3049}}
\affiliation{Algorithmiq Ltd, Kanavakatu 3C 00160 Helsinki, Finland}

\author{Sergei Filippov\,\orcidlink{0000-0001-6414-2137}}
\affiliation{Algorithmiq Ltd, Kanavakatu 3C 00160 Helsinki, Finland}

\author{Matteo A. C. Rossi\,\orcidlink{0000-0003-4665-9284}}
\affiliation{Algorithmiq Ltd, Kanavakatu 3C 00160 Helsinki, Finland}

\author{Guillermo García-Pérez\,\orcidlink{0000-0002-9006-060X}}
\affiliation{Algorithmiq Ltd, Kanavakatu 3C 00160 Helsinki, Finland}

\date{\today}

\begin{abstract}
Characterization of noise in current near-term quantum devices is of paramount importance to fully use their computational power. However, direct quantum process tomography becomes unfeasible for systems composed of tens of qubits. A promising alternative method based on tensor networks was recently proposed [Nat Commun 14, 2858 (2023)]. In this work, we adapt it for the characterization of noise channels on near-term quantum computers and investigate its performance thoroughly. In particular, we show how experimentally feasible tomographic samples are sufficient to accurately characterize realistic correlated noise models affecting individual layers of quantum circuits, and study its performance on systems composed of up to 20 qubits. Furthermore, we combine this noise characterization method with a recently proposed noise-aware tensor network error mitigation protocol for correcting outcomes in noisy circuits, resulting accurate estimations even on deep circuit instances. This positions the tensor-network-based noise characterization protocol as a valuable tool for practical error characterization and mitigation in the near-term quantum computing era.
\end{abstract}

\maketitle

\section{Introduction}

\begin{figure*}
    \centering
    \includegraphics[width=\textwidth]{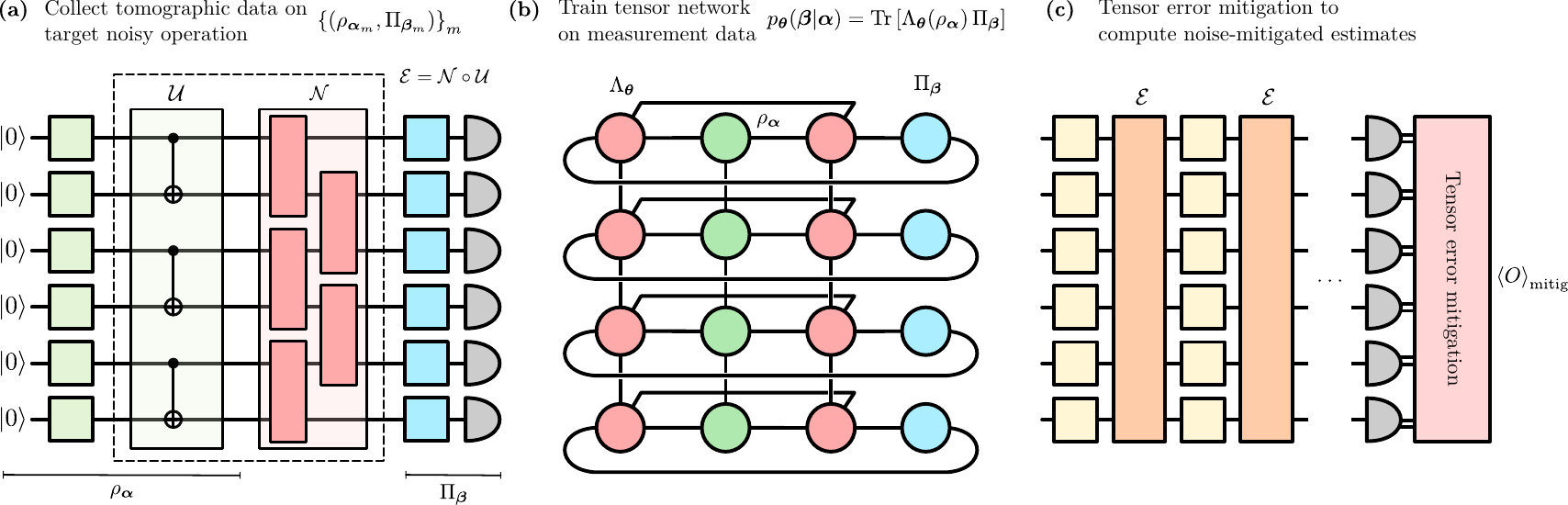}
    \caption{\textbf{(a)} Description of the tensor network-based noise characterization pipeline. The goal is to characterize the noise map $\mathcal{N}$ unavoidably accompanying an ideal unitary operation $\mathcal{U}$ (in the plot, a layer of CNOT gates) when this is executed on real quantum hardware. A number of experimental tomographic samples obtained with random preparations and measurements are collected on the noisy quantum computer to learn the noise map. For a single experimental shot, the noise channel $\mathcal{N}$ we aim to reconstruct (red rectangles) acts on the state that we denote by tomographic state $\rho_{\bm{\alpha}}$, prepared by the single-qubit gates (green squares) followed by the unitary channel $\mathcal{U}$. The state is then measured through a collection of POVMs with effects $\Pi_{\bm{\beta}}$ (blue squares). \textbf{(b)} Representation of the tomographic experiment as tensor network, where the noise channel under investigation is written as a locally-purified density operator (LPDO) $\Lambda_{\bm{\theta}}$ parameterized by the quantities $\bm{\theta}$. The noise channel is learned by training the LPDO according to a suitable cost function, so that it best explains the tomographic measurement statistics observed on the quantum device. \textbf{(c)} Tensor-network error mitigation (TEM) applied to the full noisy process $\mathcal{E}$ using the results of the noise characterization experiment.}
    \label{fig:main-figure}
\end{figure*}

Near-term quantum computers are currently entering what has been termed as the ``utility era''~\cite{KimIBMUtility2023}. This key advancement has been possible thanks to quantum error mitigation~\cite{Cai2022}, that is, a collection of techniques for mitigating and eventually eliminating the effects of noise in quantum computers without relying on quantum error correction. Despite encouraging recent progresses~\cite{Bluvstein2023}, quantum error correction is currently out of reach for useful quantum computing.

Many error mitigation techniques require (ideally perfect) knowledge of the noise channels on the device, that is, of the actual physical processes implemented on the real machine instead of the ideal unitary gates or ideal sharp measurements. Characterizing quantum processes for tens to hundreds of qubits is however not a trivial task. Standard state and process tomography~\cite{DAriano2001} requires an exponential amount of resources as a function of the number of qubits~\cite{nielsen_chuang_2010}. Different techniques have been proposed to overcome this key issue and other inherent difficulties in tomographic methods, such as the enforcement of physical constraints. Examples include, but are not limited to, twirling methods that tailor the investigated processes to specific simpler forms~\cite{Erhard2019,vanDerBerg2023}, classical shadow methods to reconstruct quantum processes~\cite{Kunjummen2023, Levy2023ShadowQPT}, tensor network methods to characterize non-Markovian evolution processes~\cite{White2023unifying, WhiteNonmarkovianPTT2022}, compressed sensing techniques for low-rank quantum processes~\cite{RodionovCompressedQPT2014, FlammiaCompressedQPT2012}, and methods that ensure meaningful tomography by appropriately restricting the reconstructed process to physical subspaces, for instance by means of constrained gradient-descent~\cite{ShahnawazGDQPT2023} or by projection onto the set of processes~\cite{SurawyStepney2022projectedleast}.

In this paper, we develop the recently introduced tensor-network-based quantum process tomography method proposed by Torlai \textit{et al.}~\cite{Torlai2023}, and apply it to the problem of noise characterization in near-term devices. This method consists in finding an efficient tensor network representation \cite{SchollwckDMRG2011,orus2014,BridgemanHandwavyTN2017,MontangeroTNBook2018} of the quantum process under scrutiny. Torlai \textit{et al.}~benchmark their method with ideal circuits (i.e unitary transformations) up to 10 qubits, and a noisy operation on 5 qubits subject to single-qubit errors. However, some applications, such as probabilistic error cancellation (PEC)~\cite{Temme2017,Endo2018} or tensor network error mitigation (TEM)~\cite{Filippov2023}, require the knowledge of the performance of individual gates or layers of gates. Therefore, here we focus on the characterization of each individual layer of a given circuit. Moreover, we exploit the fact that the process can be split into an ideal and a noisy part, and characterize only the latter. These two modifications have the advantage of alleviating the numerical requirements of the method. Finally, we consider realistic types of correlated noise, including the noisy model observed on IBM devices, and investigate systems of up to 20 qubits in size.

We study the tensor-network-based noise learning procedure by running several numerical experiments for the characterization of various correlated noise channels with brickwork-like structure and realistic noise parameters, which are of great relevance in near-term quantum computing. \mar{While our analysis is restricted to tensor networks with 1D connectivity (such as MPOs), the proposed method could be generalized to more complex topologies of qubit connectivity, albeit at the cost of more demanding classical computations.} 

In this work, we discuss the necessary amount of experimental settings and measurement shots to obtain an accurate reconstruction, and find that collecting statistics on just a limited number of random experiments with informationally complete states and measurements provides sufficient data to accomplish the task. In particular, we observe that linearly many experimental samples in the number of qubits suffice to ensure very good reconstructions. Clearly, one can improve the reconstruction by increasing the number of experimental settings (input states and measurement bases) or the number of allocated shots per setting. For example, only $10^3$ different experiments with $10^3$ measurement shots each are sufficient to characterize a correlated brickwork layer of depolarizing noise channel on $n=20$ qubits with an error of $\approx 10^{-4}$, as measured in terms of Frobenius distance between the ideal and reconstructed channels. We also confirm the good reconstruction accuracy by comparing values in the Pauli transfer matrix of the true and the reconstructed processes, and again find a good agreement.
\mar{Additionally, we address the effect of state preparation and measurement (SPAM) errors on reconstruction accuracy, demonstrating not only its robustness against small errors but also that SPAM error-free performance can be achieved by calibrating the quantum device using existing quantum detector tomography methods~\cite{Cattaneo2023}.}

We also investigate the performance of the method in conjunction with the tensor network error mitigation (TEM) protocol~\cite{Filippov2023} in noisy Clifford circuits of up to 10 qubits and 30 layers. The combined characterization and mitigation approach is capable of mitigating noise and predicting the expected value of heavy Pauli observables with high accuracy (relative error of the order of $10^{-2}$). This suggests that the characterization protocol is a valuable tool for practical error mitigation in the near-term era. In Fig.~\ref{fig:main-figure} we summarize the main idea of the presented analysis. 

The paper is structured as follows. In Sec.~\ref{sec:Protocol}, we review the tensor network representation of processes, along with Torlai \textit{et al.}'s process tomography method (with some technical modifications). In Sec.~\ref{sec:tomography_experiments}, we introduce the numerical experiments used to test the method, and describe the noise models that we have taken into consideration. The results of these numerical experiments are then presented in Sec.~\ref{sec:results}, while Sec.~\ref{sec:errorMit} is devoted to analyzing the combination of the process characterization method with the recently proposed tensor network-based error mitigation protocol (TEM). Finally, we offer some concluding remarks in Sec.~\ref{sec:conclusions}.

\section{Tensor network procedure for noise characterization}
In this section we introduce all the necessary tools for describing the tensor network noise characterization protocol, graphically summarized in Fig.~\ref{fig:main-figure}(a, b).
\label{sec:Protocol}
\subsection{Tensor network representation of noise}
\label{sec:tensorNetworkFormalism}
\begin{figure}[ht]
    \centering
    \includegraphics[width=0.45\textwidth]{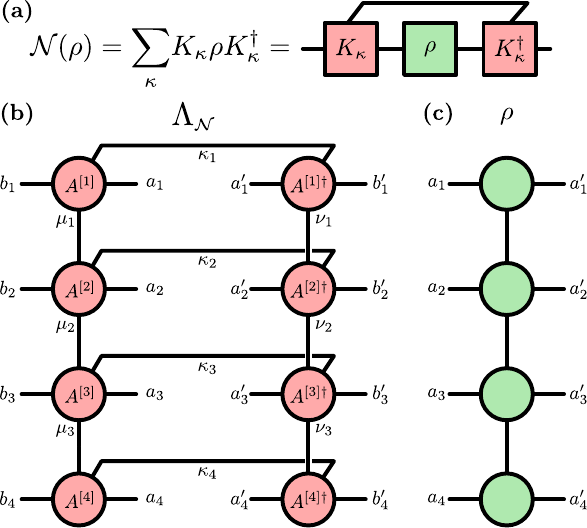}
    \caption{\textbf{(a)} Kraus representation of the noise channel $\mathcal{N}$ applied on a state $\rho$. \textbf{(b)} Tensor network representation of the channel $\mathcal{N}$ as the locally-purified density operator (LPDO) $\Lambda_\mathcal{N}$. \textbf{(c)} Tensor network representation of the state $\rho$ as matrix product operator (MPO). The action of $\mathcal{N}$ on $\rho$ in tensor network notation is obtained by connecting the two tensor networks according to the indices as in the figure.}
    \label{fig:TNstructure}
\end{figure}

Let us consider a system of $n$ qubits, the total Hilbert space $\mathcal{H}$ of which has dimension $2^n$. 
We aim at estimating a generic noise channel, which is formally described by the completely positive and trace-preserving (CPTP) quantum map $\mathcal{N}$~\cite{Watrous2018} belonging to the space of bounded operators acting on the set of density matrices of the qubits. 

Different representations of $\mathcal{N}$ are available~\cite{Watrous2018, Wilde2017, nielsen_chuang_2010}, such as the Choi matrix or the Liouville superoperator representation~\cite{WoodTNGraphical2015}. For our purposes, we choose to represent $\mathcal{N}$ through its \textit{Kraus decomposition}, defined implicitly by
\begin{equation}
    \label{eqn:krausDec}
    \mathcal{N}[\rho] = \sum_{\kappa} K_\kappa\,\rho\,K_\kappa^\dagger\,, 
\end{equation}
where $\rho$ is a quantum state, and $K_\kappa$ are \textit{Kraus operators} acting on the Hilbert space of the qubits, satisfying the trace preserving condition
\begin{equation}
\label{eqn:KrausNormalization}
    \sum_\kappa K_\kappa^\dagger K_\kappa = \mathbb{I}\,.
\end{equation} 
For any $n$-qubit map $\mathcal{N}$, the Kraus operators can be chosen in such a way that their number is at most $4^n$. 

Each representation of $\mathcal{N}$ can be described in different ways using the tensor network formalism~\cite{WoodTNGraphical2015}. In our case, we focus on a tensor network representation of the Kraus operators known in the literature as \textit{locally-purified density operator} (LPDO)~\cite{Werner2016,Guo2022,Torlai2023},
which is depicted in Fig.~\ref{fig:TNstructure}. We denote the LPDO representation of a channel $\mathcal{N}$ as  $\Lambda_\mathcal{N}$, and we refer to Appendix~\ref{sec:LPDO_definition} for the explicit expression of the tensor components of such tensor network.

Let $\rho \in \mathbb{C}^{2^n \times 2^n}$ be the density matrix of a system of $n$ qubits. Such matrix can be written in Matrix Product Operator (MPO) form as~\cite{VerstraeteMPO2004, WoodTNGraphical2015, MontangeroTNBook2018, BridgemanHandwavyTN2017, SchollwckDMRG2011}
\begin{equation}
    \label{eq:rho_mpo}
    \rho = \sum_{\lambda_1, \ldots, \lambda_{n-1}} \rho^{[1]}_{\lambda_1} \otimes \rho^{[2]}_{\lambda_1, \lambda_2} \otimes \ldots \otimes \rho^{[n]}_{\lambda_{n-1}}\,.
\end{equation}
Then, using the LPDO representation for the quantum channel $\Lambda_{\mathcal{N}}$, and the MPO form for a state $\rho$, the action of the quantum channel $\mathcal{N}[\rho]$ can be written in tensor network notation as
\begin{align}
    \label{eqn:LPDOaction}
    \Lambda_{\mathcal{N}}[\rho] = \sum_{\bm{\kappa}=1}^{\chi_\kappa}~& \sum_{\bm{\mu}=1}^\chi\qty(A_{\mu_1,\kappa_1}^{[1]}\otimes A_{\mu_{1},\mu_2,\kappa_2}^{[2]}\otimes\ldots\otimes  A_{\mu_{n-1},\kappa_n}^{[n]}) \nonumber \\ 
    &\quad\quad\quad \sum_{\bm{\lambda}} \rho^{[1]}_{\lambda_1} \otimes \rho^{[2]}_{\lambda_1, \lambda_2} \otimes \ldots \otimes \rho^{[n]}_{\lambda_{n-1}}\\
    & \sum_{\bm{\nu}=1}^\chi \qty(A^{[1]\dagger}_{\nu_1,\kappa_1} \otimes A_{\nu_{1},\nu_2,\kappa_2}^{[2]\dagger} \otimes \ldots \otimes  A_{\nu_{n-1},\kappa_n}^{[n]\dagger})\,,\nonumber
\end{align}
where $\smash{A^{[m]}_{i,j,k} \in \mathbb{C}^{2\times2}}$ are local operators acting on single-qubit sites, $\smash{A^{[m]\dagger}_{i,j,k} \coloneqq (A^{[m]}_{i,j,k})^\dagger}$ are their transposed conjugates, and they interact with the corresponding local matrices of $\rho$ as shown in Fig.~\ref{fig:TNstructure}. The indices $\bm{\mu}=(\mu_1, \ldots, \mu_{n-1})$ and $\bm{\nu}=(\nu_1, \ldots, \nu_{n-1})$ with $\smash{\mu_j, \nu_j = 1, \ldots, \chi_b^{(j)}}$ are the so-called virtual \textit{bond indices} of the LPDO, while $\bm{\kappa}=(\kappa_1, \ldots, \kappa_n)$ with $\smash{\kappa_j = 1, \ldots, \chi_\kappa^{(j)}}$ are called \textit{Kraus indices}. The maximal \textit{bond dimension} of the LPDO is defined as the size of the largest of the virtual bond indices, $\chi_b =\max_{j}\chi_b^{(j)}$. Similarly, the maximal \textit{Kraus dimension} is instead given by $\smash{\chi_{\kappa} = \max_{j}\chi_\kappa^{(j)}}$. 
The LPDO structure has also been used to represent the positive Choi matrix of the channel $\mathcal{N}$ in a very similar way~\cite{Torlai2023, Guo2022}. Finally, the Kraus decomposition expressed by the LPDO can be easily transformed into an MPO which is the superoperator representation of the quantum channel in Liouville space~\cite{WoodTNGraphical2015}, as described in Appendix~\ref{app:LPDO_to_MPO}.

It is important to stress that we consider the task of characterizing shallow processes with a clear local structure, which admit an efficient classical tensor-network representation with low bond dimensions. \mar{By \textit{shallow}, we hereby refer to quantum operations acting on $n$ qubits that create only short-range qubit-qubit correlations, with this range being independent of $n$. For instance, a layer of CNOTs is shallow because it creates entanglement between pairs of qubits only, so the range of the correlations is 2, independently of the total number of qubits. Our method may still be used to characterize noise in more complex circuits but, in order to avoid an exponential scaling of the bond dimension of their tensor network representations, each deep quantum circuit should be divided into elementary shallow layers, with our protocol applied independently to each of them.}

Applying the LPDO on the MPO $\rho$ in Eq.~\eqref{eqn:LPDOaction} corresponds to the action of the Kraus decomposition of the channel $\mathcal{N}$ on a quantum state. Indeed, we can group the Kraus indices in a single multi-index $\kappa = \{\kappa_1,\ldots,\kappa_{n}\}$, the upper summation limit of which is equal to the product of all the dimensions of the Kraus indices. Then, we obtain a single set of global Kraus operators $K_\kappa$, the MPO structure of which is given by 
\begin{equation}
\label{eqn:KrausDecTN}
    K_\kappa =\sum_{\mu_1,\ldots,\mu_{n-1}} A_{\mu_1,\kappa}^{[1]}\otimes A_{\mu_{1},\mu_2,\kappa}^{[2]}\otimes\ldots\otimes  A_{\mu_{n-1},\kappa}^{[n]},
\end{equation}
and the action of the channel $\mathcal{N}$ on the state $\rho$ in Eq.~\eqref{eqn:LPDOaction} is equivalent to Eq.~\eqref{eqn:krausDec}.

Note, however, that while the LPDO structure is completely positive by design, it does not automatically satisfy the trace preserving condition of Eq.~\eqref{eqn:KrausNormalization}, which translates to $\text{Tr}_{\bm{b}, \bm{b}'}[\Lambda_{\bm{\theta}}]=\mathbb{I}$ in LPDO notation~\cite{WoodTNGraphical2015}, where $\Lambda_{\bm{\theta}}$ is a generic LPDO parameterized by quantities $\bm{\theta}$, and the indices $\bm{b}$ and $\bm{b}'$ in the partial trace are as in Fig.~\ref{fig:TNstructure}. As proposed in~\cite{Torlai2023}, one can enforce such property by adding a trace preservation penalty term in the cost function at training time, but more explicit constraints have also been proposed~\cite{SrinivasanTPLPDO2021}. In addition, as discussed in Sec.~\ref{sec:TNoptimization}, one can initialize the tensor elements $\bm{\theta}$ in such a way that the resulting LPDO is at least correctly normalized $\Tr[\Lambda_{\bm{\theta}}] \approx 2^n$ in expectation value.

\subsection{Data sampling}
\label{sec:dataSampling}
In what follows, we describe the tomographic sampling strategy based on the recent work by Torlai \textit{et al.}~\cite{Torlai2023}. It consists in generating $N_\text{set}$ experimental tomographic settings in a randomized way in order to obtain sufficient information about the quantum channel we aim to characterize. For experiments on near-term quantum computers, we define an experimental tomographic setting as a collection of $n$ separable, single-qubit input states, and a collection of $n$ choices of measurement basis, one for each qubit of the device. 

In our work, for each qubit, we only use three possible measurement bases, corresponding to the Pauli measurements, i.e. we measure along the $X$, $Y$, or $Z$ axes. The POVM effects of the measurement along the $X$ axis, for instance, are $\ket{x}\!\bra{x}$ and $\ket{-x}\!\bra{-x}$, where $\ket{x}$ ($\ket{-x}$) is the eigenstate of $X$ with eigenvalue $+1$ ($-1$), and equivalently for other axes. Since the typical native measurement on a quantum computer is in the computational basis, we can perform a measurements in the $X$ or $Y$ directions by simply applying suitable single-qubit rotations before the measurement.

Each single-qubit input state, which for the $j$-th qubit is denoted by $\rho_{\alpha_j}^\text{in}$, is drawn from an informationally complete (IC) pool of states $\mathcal{P}=\{\rho_{\alpha_j}^\text{in}\}_j$, forming a basis in the space of single-qubit density matrices. The minimum number of states in the pool is 4, but this set may also be larger. For instance, one may choose the overcomplete set comprising the 6 Pauli eigenstates $\mathcal{P}=\{\ket{\pm x}\!\bra{\pm x},\ket{\pm y}\!\bra{\pm y},\ket{\pm z}\!\bra{\pm z}\}$, which can be prepared by applying a single-qubit unitary gates to state $\ket{+z}=\ket{0}$. \mar{Alternatively, for the sake of employing less experimental settings, one may also use a symmetric and informationally complete (SIC) set of input states containing only four elements, for example given by~\cite{vrehavcek2004minimal}
\begin{equation}
    \label{eqn:SICstates}
\begin{split}
    \rho_j &= \frac{1}{4}(1+ \bm{a}_j \cdot \bm{\sigma})\,,\quad \text{with} \\
    \bm{\sigma} &= (X, Y, Z)\,,\\
    \bm{a}_0 &= \frac{(+1,+1,+1)}{\sqrt{3}}\,,~\bm{a}_1 = \frac{(+1,-1,-1)}{\sqrt{3}}\,, \\
    \bm{a}_2 &= \frac{(-1,+1,-1)}{\sqrt{3}}\,,~ 
    \bm{a}_3 = \frac{(-1,-1,+1)}{\sqrt{3}}\,.
\end{split}
\end{equation}}
In practical scenarios, one usually seeks to characterize the noise channel $\mathcal{N}$ accompanying an ideal unitary layer $\mathcal{U}$. Thus, we define the \textit{tomographic state} as $\rho_{\boldsymbol{\alpha}} = \mathcal{U}[\rho_{\boldsymbol{\alpha}}^\text{in}]$, obtained by evolving the initial randomly chosen input states through the ideal unitary whose noise we want to characterize, see Fig.~\ref{fig:main-figure}(a).

An experimental setting is then defined in the following way. We first draw one state $\rho_{\alpha_j}^\text{in}$ for each qubit $j=1,\ldots,n$ from a uniform distribution over the pool $\mathcal{P}$, which, for simplicity, we assume to be equal for all qubits. Let us use the collective index $\boldsymbol{\alpha}=\{\alpha_1,\ldots,\alpha_n\}$ to denote the choice of initial states for all the qubits. Next, we draw one measurement basis $\beta_j \in \{X,Y,Z\}$ for each qubit $j$, again from a uniform distribution, and use the collective index $\boldsymbol{\beta} = \{\beta_1,\ldots,\beta_n\}$ to indicate which measurement basis has been chosen on each qubit. 

A single tomographic experiment will then consist in preparing the state $\rho_{\boldsymbol{\alpha}}^\text{in}=\bigotimes_{j=1}^n\rho_{\alpha_j}^\text{in}$ on all qubits, evolving it through the \textit{full noisy process} $\mathcal{E} = \mathcal{N}\circ\,\mathcal{U}$, and finally measuring each qubit in the proper basis $\beta_j$. Note that, since we know the logical operation $\mathcal{U}$ the noise channel of which we are characterizing, we isolate the noise channel $\mathcal{N}$ from the full noisy process, and regard this experiment as the application of an unknown channel $\mathcal{N}$ onto the known tomographic state $\rho_{\boldsymbol{\alpha}} = \mathcal{U}[\rho_{\boldsymbol{\alpha}}^\text{in}]$. Note that the tomographic state is in general entangled, but its spatial correlations (which impacts the bond dimension of its MPO representation) are short range if the unitary circuit $\mathcal{U}$ is shallow. This is the case we are interested in, as we are considering noise affecting single-layer instructions. This means that we can represent $\rho_{\bm{\alpha}}$ as a tensor network efficiently.

A single-qubit Pauli measurement can be described by a POVM with only two effects corresponding to outcome $\zeta=+1$ or $\zeta=-1$, so the outcome of a single $n$-qubit tomographic experiment can then be represented as a vector $\boldsymbol{\zeta}=(\zeta_1,\ldots,\zeta_n)$. The probability of obtaining the outcome $\boldsymbol{\zeta}$ for a fixed experimental setting, defined by the choice of input states $\bm{\alpha}$ and measurement bases $\bm{\beta}$, is given by the Born rule
\begin{equation}
    \label{eq:transition_probs}
    p(\boldsymbol{\zeta}|\boldsymbol{\alpha},\boldsymbol{\beta}) = \Tr[\mathcal{N}[\rho_{\boldsymbol{\alpha}}]~\bigotimes_{j=1}^n \Pi_{\zeta_j}(\beta_j)],
\end{equation}
where $\Pi_{\zeta_j}(\beta_j)$ is the effect corresponding to the outcome $\zeta_j$ for a measurement in the $\beta_j$ basis performed on the $j$-th qubit.

We point out that this sampling strategy assumes that we know the input states and the measurements perfectly well. It is however well known that this is not the case on near-term quantum computers, where state preparation and measurement (SPAM) errors are currently unavoidable~\cite{Merkel2013}. This consideration has led to different self-consistent tomographic methods to determine the input states, the computational gates, and the measurement outcomes consistently and simultaneously~\cite{Merkel2013, Nielsen2021}. Unfortunately, these procedures are usually too resource-expensive for useful near-term applications (i.e., involving tens of qubits), even when considering optimized strategies~\cite{Brieger2023} \mar{(a more promising procedure is considered in Ref.~\cite{Helsen2023}, which performs SPAM-robust shadow estimation of some properties of a gate set from random gate sequences)}. For this reason, in the case of tensor network noise characterization, one may adopt a more practical solution on a real quantum computer: before running the noise tomography experiment, one can perform a calibration of the machine yielding a self-consistent description of the input states and measurements, such as the one based on semidefinite programming recently proposed by some of the authors~\cite{Cattaneo2023}. Then, the output of the protocol, that is, a set of input states and POVM effects that are self-consistent and capture what is physically prepared and measured on the device, would be used in the noise characterization procedure, as in Eq.~\eqref{eq:transition_probs}. \mar{We point out that all the results of this paper, apart from the ones discussed in Sec.~\ref{ssec:spam-test}, do no take SPAM errors into account and do not employ self-consistent tomography. We leave a detailed study of the efficacy of self-consistent strategies to ameliorate SPAM errors in the tensor network noise characterization protocol for future works.}

\subsubsection{Efficiency of the sampling strategy} 
Let $R=\abs{\mathcal{P}}$ be the number of possible input states in the pool $\mathcal{P}$, for which we know that $R \geq 4$. For each qubit we then have a total of $3R$ possible different experimental settings, given by all the possible combinations of input states ($R$) and measurement bases (the $3$ Pauli measurements). For a system of $n$ qubits and assuming independent preparations and measurements, the total number of different settings therefore is equal to $(3R)^n$, which is a formidable number even for $n \sim 10$. Such exponential scaling, typical of quantum tomography, makes it unfeasible to implement all the possible different settings in a tomographic experiment when a large number of qubits is used.

We circumvent the issue by randomly generating only $N_\text{set}$ different tomographic settings using the procedure described above, and allocating a number of $N_\text{shots}$ measurement shots to each of these settings. Therefore, a complete tomographic experiment will use a total of $N = N_\text{shots} \times N_\text{set}$ measurements. Despite not having the exponentially large tomographic data required to reconstruct arbitrary channels, such limited information can still be sufficient in our scenario, where we are interested in learning processes with a local structure, which can be effectively described using a only limited number of values. Additionally, as already stated above, the local structure of the noise is also a key assumption for its efficient description through a LPDO tensor network with a small bond dimension.

We point out that here our analysis deviates from that performed in Ref.~\cite{Torlai2023}, where the effect of number of shots per setting is not taken into account. Instead, in this work we consider a scenario that is more realistic for near-term quantum computers, and in particular for superconducting quantum devices available on the cloud~\cite{ibmQ}, where the total measurement budget $N$ is allocated by executing $N_\text{shots}$ shots on each of a limited number of experimental settings $N_\text{set}$. In fact, given access constraints and the relatively long wall-time needed to compile instructions on current near-term hardware~\cite{ibmQ, Qiskit}, it is of paramount importance to be able to extract relevant information out of only a limited number of distinct experimental setups. The number of settings $N_\text{set}$ is often the practical bottleneck for current experiments, while the number of shots per settings $N_\text{set}$ comes at a much lower cost, both in terms of accessibility and execution time. 

We point out that such random sampling strategy, based on generating only a reduced number of settings that in principle is not sufficient for full process tomography of the quantum channel, is equivalent to sampling for \textit{shadow tomography}~\cite{Aaronson2020,Huang2020,Acharya2021} of quantum processes, which has been explored in some recent works~\cite{Kunjummen2023,Levy2023ShadowQPT,Acharya2023,Helsen2023}. The difference between our method and shadow process tomography lies in the post-processing of the sampled data. Instead of applying linear inversion~\cite{Kunjummen2023} or using more refined fitting methods~\cite{Levy2023ShadowQPT,Acharya2023,Helsen2023} starting from the raw data, we train the LPDO structure to obtain the most accurate tensor network description of the channel, \mar{by finding the parameters in the tensor network that best explain the experimental data} see Sec.~\ref{sec:TNoptimization}. We leave a direct comparison of our approach with those based on shadow tomography as a topic for future studies.

\subsubsection{Alternative local sampling strategies} 
The sampling strategy described so far generates different global random settings for the tomographic experiment. However, one may also adopt a different strategy, which assumes that the process $\mathcal{N}$ under investigation only generates local correlations, and may therefore be well-characterized by using only local information about the subsystems. Broadly speaking, such methods select the tomographic settings in a way that one is able to collect data on all reduced subsystems of given locality, and then reconstruct the whole channel based on such local information. Local tomographic strategies building on such ideas have been successfully used in the literature to characterize quantum states with local correlations~\cite{Guo2023, Cramer2010, Lanyon2017}.

However, such strategies cannot be straightforwardly applied to the case of process tomography, where one has to probe the channel under investigation not only with informationally complete measurements, but also input states (we don't consider the reduction of channel to state tomography via the Choi–Jamiołkowski isomorphism as this requires the use of ancillary qubits~\cite{AltepeterAAQPT2003}). Taking into account the burden of state generation, one can check that the resources needed to collect local tomographic data quickly become experimentally unfeasible, even for low locality. We refer the interested reader to Appendix~\ref{sec:localStrategy}, where we discuss in detail possible tomographic strategies for accessing local data, also based on lightcone arguments stemming from the brickwork structure of the channel under investigation. In addition, in appendix~\ref{app:local_vs_global}, we also provide preliminary numerical evidence that the global random strategy performs better than a simple local strategy, both in terms of total measurements needed and reconstruction accuracy. All the numerical results presented in the following are thus obtained following the random generation of tomographic settings described in Sec.~\ref{sec:dataSampling}.

\subsection{Tensor network optimization \label{sec:TNoptimization}}
The optimization of the LPDO $\Lambda_{\bm{\theta}}$ over a set of parameters $\bm{\theta}$ is based on the approach proposed in Ref.~\cite{Torlai2023}, in which the tensor network is trained so that the predicted distribution of outcomes best matches the observed measurement statistics. 

Formally, for $N$ total experimental shots, let $S= \{(\rho_{\bm{\alpha}_m},\, \Pi_{\bm{\zeta}_m}(\bm{\beta}_m)\}_{m=1}^N$ be the tomographic dataset collected on a real quantum device consisting of $N$ pairs of tomographic states $\rho_{\bm{\alpha}_m}$ and corresponding measured effects $\Pi_{\bm{\zeta}_m}(\bm{\beta}_m)$, where subscript $m$ labels single experimental shots. Then, the LPDO can be fitted to the experimental data by minimizing the objective function~\cite{Torlai2023}
\begin{equation}
\begin{split}
\label{eq:kl_divergence}
    D_{\scriptscriptstyle \text{KL}}(\bm{\theta}; S) & = - \frac{1}{N} \sum_{m=1}^N \log p(\bm{\zeta}_m|\bm{\alpha}_m,\bm{\beta}_m;\bm{\theta}) \\
    & = - \frac{1}{N} \sum_{m=1}^N \log \Tr[\Lambda_{\bm{\theta}}(\rho_{\bm{\alpha}_m})\,\Pi_{\bm{\zeta}_m}(\bm{\beta}_m)]\,,
\end{split}
\end{equation}
which is a Monte Carlo approximation to the Kullback-Leibler divergence between the true probability distribution of the quantum process under investigation~\eqref{eq:transition_probs}, and the one generated by the parameterized tensor network.

In addition, the authors in Ref.~\cite{Torlai2023} propose to using an additional penalty term in the loss function that favors physically valid LPDO satisfying the trace preservation (TP) condition~\eqref{eqn:KrausNormalization}. Such penalty term is given by the normalized Frobenius distance between the identity and the MPO obtained by contracting the outer legs in the LPDO ($\bm{b}$ and $\bm{b}'$ in Fig.~\ref{fig:TNstructure}), namely
\begin{equation}
    \label{eq:trace_penalty}
    \delta_{\scriptscriptstyle \text{TP}}(\bm{\theta}) = \dfrac{\norm{\Tr_{\scriptscriptstyle \text{out}}\qty[\Lambda_{\bm{\theta}}] - \mathbb{I}}_F}{2^{n/2}}\,,
\end{equation}
where $\smash{\norm{A}_F^2 \coloneqq \Tr[A^\dagger A]}$ is the operator Frobenius norm.

Finally, the complete loss function used to drive learning process is then given by
\begin{equation}
    \label{eq:loss_function}
    L(\bm{\theta};\, S) = D_{\scriptscriptstyle \text{KL}}(\bm{\theta};\,S) + \eta~\delta_{\scriptscriptstyle \text{TP}}(\bm{\theta})\,,
\end{equation}
where $\eta \in \mathbb{R}$ is a hyperparameter tuning the importance of the TP condition in the training process. In all our numerical simulations we set $\eta = 1.2$, which was \mar{heuristically} found to consistently ensure a good convergence to a properly normalized LPDO, $\Tr[\Lambda_{\bm{\theta}}] \approx 2^n$, at the end of training. \mar{While different choices don't impact the end results sensibly, these may result either in slower convergence times towards physically meaningful solutions, or to solutions having incorrect ---but still close, if $\eta \approx 1$--- trace. Also, we note that such hyperparameter could be itself adapted during training, but we leave this investigation as a topic for future studies.}

Despite its effectiveness in training the LPDO, the loss function~\eqref{eq:loss_function} does not satisfy the symmetry requirement for a true distance and has a limited physical interpretation. For this reason, we measure the reconstruction error in the characterization procedure through the quantity
\begin{equation}
    \label{eq:characterisation_error}
    \Delta(\Lambda,\,\Lambda_{\bm{\theta}}) = \dfrac{\norm{\Lambda - \Lambda_{\bm{\theta}}}_F^2}{2^{2n}}\,,
\end{equation}
consisting of a properly normalized Frobenius distance between the true channel $\Lambda$ and the trainable one $\Lambda_{\bm{\theta}}$, similar to the fidelity-like error measure used in~\cite{Torlai2023}. Needless to say, this measure is only available in classical numerical simulations, where the true channel is known. 

\subsubsection{Normalized initialization of the LPDO \label{ssec:lpdo_initializatio}}
At the start of the training procedure, the parameterized LPDO is initialized with random values. However, this typically leads to unphysical quantum maps not respecting either the TP constraint~\eqref{eq:trace_penalty} or the normalization condition $\Tr[\Lambda] = 2^n$. In order to alleviate this issue, we first employ a parameter initialization method that yields, in expectation value, a correctly normalized LPDO, and then variationally pre-optimize the tensor network in order to appropriately satisfy the trace preserving condition. Both these strategies were heuristically found to improve convergence to good solutions and to stabilize the training process by avoiding numerical instabilities related to unphysical initializations of the tensor network.

The tensor elements $\theta_k \in \mathbb{C}$ of the LPDO $\Lambda_{\bm{\theta}}$ are randomly initialized from a complex Gaussian distribution
\begin{equation}
\label{eq:random_complex_vars_main}
\begin{split}
    & \theta_k = \Re(\theta_k) + i \Im(\theta_k) \\
    & \textrm{with}\,\, \Re(\theta_k),\, \Im(\theta_k) \sim \mathcal{G}(0, \sigma^2)\,,
\end{split}
\end{equation}
where $\mathcal{G}(0, \sigma^2)$ denotes a Gaussian distribution with zero mean and variance $\sigma^2$. As proven in Appendix~\ref{app:trace_lpdo}, under such circumstances, one can explicitly compute the expectation value of the trace of the LPDO $\Tr[\Lambda_{\bm{\theta}}]$ upon initialization, which amounts to
\begin{equation}
\label{eq:exp_gaussian_trace_lpdo_main}
    \mathbb{E}_{\bm{\theta}}[\Tr[\Lambda_{\bm{\theta}}]] = \qty(8\sigma^2\chi_\kappa)^n\, \chi_{b}^{n-1}\,,
\end{equation}
where $n$ is the number of qubits, and $\chi_\kappa$ and $\chi_b$ are the Kraus dimension and the virtual bond dimension of the LPDO, respectively. Thus, by sampling the initial parameters according to a Gaussian with variance
\begin{equation}
    \left. \sigma^2 = 2 \middle/ \qty(8 \chi_\kappa \chi_b^{1 - 1 / n})\right.
\end{equation}
the LPDO is properly normalized to the correct value $\mathbb{E}_{\bm{\theta}}[\Tr[\Lambda_{\bm{\theta}}]] = 2^n$ on average upon random initialization.

Additionally, we further pre-optimize the initial LPDO to satisfy the TP constraint by variationally minimizing the penalty term $\delta_{\scriptscriptstyle \text{TP}}(\bm{\theta})$~\eqref{eq:trace_penalty} by means of an optimizer before the actual training of the LPDO starts.

\subsubsection{Details on numerical simulations and optimization}
All numerical experiments are run using the python tensor network library \texttt{quimb}~\cite{GrayQuimb2018}, in combination with libraries for automatic differentiation and optimization \texttt{jax}~\cite{jax2018github} and \texttt{optax}~\cite{OptaxDeepmind2020jax}. 

The trace preserving pre-optimization of the LPDO is run using optimizer L-BFGS-B~\cite{L-BFGS-B_Optimizer} provided in \texttt{quimb}. The training of the LPDO by minimization of the loss function $L(\bm{\theta};\, S)$~\eqref{eq:loss_function} is done using the Adam optimizer~\cite{Kingma2017adam}, together with an additional custom exponential decay schedule of the learning rate, which was found to improve convergence. We refer to Appendix~\ref{app:training_details} for further details on the optimization process, including details on the training batch size and dimension of the test set. 

\section{Noise tomography experiments\label{sec:tomography_experiments}}
For the sake of benchmarking our noise characterization method, we consider the task of determining the noise $\mathcal{N}$ accompanying the simple yet very common logical instruction $\mathcal{U}$ consisting of an $n$-qubit even layer of CNOTs, as depicted in Fig.~\ref{fig:main-figure}(a). We simulate different noise models applied to this circuit layer, which also take into account crosstalk errors between nearby qubits. We run some classical simulations of the tomographic experiment to characterize such a noisy circuit, and we compare the results of the tensor network reconstruction with the true noisy channel. 

In the current section, we describe the different noise models we have employed in the classical simulation. The ensuing Section~\ref{sec:results} is devoted to the study of the numerical results and performance analysis. Finally, in Sec.~\ref{sec:errorMit}, we also validate the accuracy of the channel characterization scheme by employing it to mitigate the noise on a noisy circuit through the recently proposed tensor network error mitigation protocol (TEM)~\cite{Filippov2023}.

In our simulations, we choose three different realistic noise models that are of particular importance for near-term quantum computers: the sparse Pauli-Lindblad noise model~\cite{vanDerBerg2023}, the incoherent depolarizing noise model, and the coherent depolarizing noise model. Notably, all these multi-qubit correlated noise models can be graphically represented by the \textit{brickwork} circuit structure depicted in Fig.~\ref{fig:main-figure}.

\subsection{Sparse Pauli-Lindblad noise model}
\label{sec:pauliLindbladNoiseModel}
The sparse Pauli-Lindblad noise model is a locally correlated noise model that was recently introduced as an effective method to describe errors in superconducting quantum hardware~\cite{vanDerBerg2023}. Such noise model is described by the map
\begin{equation}
\label{eqn:PauliLindbladNoise}
    \mathcal{N}_{\scriptscriptstyle \text{SPL}}[\rho] = \prod_{k\in\mathcal{K}} \qty(\omega_k \cdot +(1-\omega_k)P_k \cdot P_k)~ \rho,
\end{equation}
where $\mathcal{K}$ is a $\text{poly}(n)$-size subset of the $4^n$ $n$-qubit Pauli operators, and $\cdot$ is a placeholder for the argument of the quantum map, e.g., $(P_k \cdot P_k) \rho = P_k \rho P_k$. The coefficients $\omega_k$ are defined as 
\begin{equation}
    \omega_k = \qty(1+e^{-2\lambda_k})/2
\end{equation}
with $\lambda_k \geq 0$ being non-negative parameters defining the strength of each Pauli interaction term in the Lindblad master equation description of the noise $\mathcal{N}_{\scriptscriptstyle \text{SPL}}[\rho] = \exp[\textsf{L}](\rho)$, generated by $\textsf{L}(\rho) = \sum_{k \in \mathcal{K}}\lambda_k ( P_k \rho P_k - \rho)$~\cite{vanDerBerg2023}.
These parameters can be estimated, for instance, by \textit{cycle benchmarking}~\cite{Erhard2019,vanDerBerg2023}, which removes any preparation and measurement (SPAM) errors~\cite{Merkel2013}. \mar{However, this technique is known to provide only an ambiguous reconstruction of the parameters~\cite{Chen2023, vanDerBerg2023}.
The sparse Pauli Lindblad noise model is typically a faithful description of the noise channels on the device if one performs randomized compiling~\cite{Wallman2016} to approximately transform the possibly coherent true noise to an incoherent Pauli channel.}

The expression of the noise in Eq.~\eqref{eqn:PauliLindbladNoise} may take into account crosstalk errors between very far away qubits, depending on how we choose the set $\mathcal{K}$. To opt for a more realistic description of spatially correlated errors on the CNOT layer, we choose $\mathcal{K}$ such that it only accounts for first-neighbors crosstalk errors. That is, the coefficients $k$ in $\mathcal{K}$ can only refer to single- or two-qubit Pauli interaction terms acting on adjacent qubits, an assumption that has been experimentally validated several times~\cite{vanDerBerg2023,KimIBMUtility2023}. Importantly, in our simulations we use realistic noise coefficients for SPL noise found in current superconducting quantum hardware (see Appendix~\ref{sec:appendixSparsePauliLindblad} for further details and explicit coefficients).

We point out that the Pauli-Lindblad channel gives rise to a \textit{Clifford} noise model \cite{nielsen_chuang_2010}, that is, the application of $\mathcal{N}_{\scriptscriptstyle \text{SPL}}$ onto a Pauli operator returns the same Pauli operator scaled by a factor. 

\subsection{Incoherent depolarizing noise model}
\label{sec:incohDep}
A two-qubit depolarizing noise channel is defined as
\begin{equation}
    \label{eqn:DepolChannelTwoQ}
    \mathcal{D}^{(p)}[\rho]=(1-p)\,\rho+\frac{p}{4}\Tr[\rho]\,\mathbb{I},
\end{equation}
where $p \in [0, 1]$ is the error rate. In the incoherent depolarizing noise model, for each CNOT gate in the unitary layer $\mathcal{U}$, we apply one two-qubit depolarizing channel with error rate $p$ on the target and control qubits. Moreover, in order to simulate first-neighbors crosstalk errors, we consider another layer of two-qubit depolarizing channels with error rate $p/2$ on the nearby qubits that are not connected by a CNOT gate. 
This then creates the \textit{brickwork} structure of Fig.~\ref{fig:main-figure}(a), as the noise model consists of one even layer of two-qubit depolarizing channels followed by an odd layer with a lower error rate. Formally, the total noise channel on $n$ qubits can be expressed as follows
\begin{equation}
\label{eq:totalDepChannel}
    \mathcal{N}_{\text{inc}}^{(p)} = \prod_{k=1}^{\lfloor (n-1)/2\rfloor} \mathcal{D}_{2k,\,2k+1}^{(p/2)} \circ \prod_{k=1}^{\lfloor n/2\rfloor} \mathcal{D}_{2k-1,\,2k}^{(p)}\,,
\end{equation}
where the depolarizing channel $\mathcal{D}_{k,\, k+1}^{(p)}$ is acting on the qubits $k$ and $k+1$. 

For the simulations in the main text we set the depolarization strength to $p=10^{-3}$, which is only slightly lower than two-qubit gate errors reported for state-of-art machines based on, e.g., superconducting circuits~\cite{ibmQ, mckay2023benchmarking}, neutral atoms~\cite{Bluvstein2023}, and ion-traps~\cite{MosesIonTrapBenchmark2023}, and foreseeably achievable in the near future. However, for completeness and as discussed in the ensuing sections, we also report results for a stronger depolarizing rate in Appendix~\ref{app:strong_depol_results}.

We note that, as for the sparse Pauli-Lindblad noise model~\eqref{eqn:PauliLindbladNoise}, also the incoherent depolarizing noise model is a Clifford map. 

\subsection{Coherent depolarizing noise model}
\label{sec:cohDep}
We extend our analysis to coherent error sources by considering the more complex case where, in addition to the brickwork depolarizing channel described above, the qubits are also affected by undesired single-qubit unitaries. Specifically, we assume that the overall noise process consists of a first layer of single-qubit random rotations used to simulate coherent noise, and the aforementioned correlated incoherent depolarizing error channel.

The complete noise channel can then be written as
\begin{equation}
    \mathcal{N}_{\text{coh}}^{(p)} = \mathcal{N}_{\text{inc}}^{(p)} \circ\, \bigotimes_{j=1}^n~U_j\,,
\end{equation}
where $U_j$ are single-qubit random rotations. Given three angle parameters $\psi$, $\varphi$ and $\phi$, these random rotations can be parameterized as
\begin{equation}
    \label{eqn:RandomRotation}
    U(\phi,\varphi,\psi)=\begin{pmatrix}
        e^{i\varphi} \cos\phi & e^{i\psi}\sin\phi\\
       -e^{-i\psi}\sin\phi & e^{-i\varphi} \cos\phi \\
    \end{pmatrix}.
\end{equation}
A single-qubit Haar-random unitary rotation can then be obtained by sampling uniformly $\psi$ and $\varphi$ from $[0,2\pi]$, and $\phi=\arcsin\sqrt{\zeta}$ where $\zeta$ is sampled uniformly from $[0,1]$~\cite{ozols2009generate}. Since we assume single-qubit errors to be small, we consider restricted rotation angles given by $\zeta \leftarrow \epsilon\,\zeta$ and $\varphi \leftarrow \epsilon\,\varphi$ with $\epsilon = 10^{-3}$.

We sample one different random rotation for each qubit. If we apply the noisy circuit layer more than once, the random rotations on each qubit are the same for all layers, that is, we always associate the same noise channel to the same logical instruction (the even CNOT layer in our cases). Contrary to the sparse Pauli-Lindblad and incoherent depolarizing noise models, the coherent depolarizing noise model can be non-Clifford.

\section{Results \label{sec:results}}
\begin{figure}[t]
    \centering
    \includegraphics[width=0.475\textwidth]{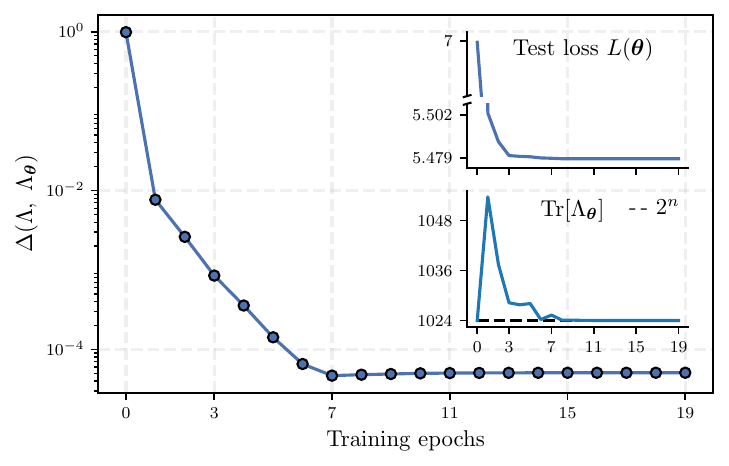}
    \caption{Example training run of the parameterized LPDO $\Lambda_{\bm{\theta}}$ for learning the brickwork depolarizing noise $\mathcal{N}^{\text{dep}}_p$ in Eq.~\eqref{eq:totalDepChannel} with $p=10^{-3}$ on a system of $n=10$ qubits. The LPDO was trained with a dataset of $10^6$ samples ($10^3$ experimental settings, $10^3$ shots per setting), and with bond dimensions $\chi_{\kappa}=2$, $\chi_b = 16$. Both the reconstruction error $\Delta(\Lambda,\,\Lambda_{\bm{\theta}})$ and the test loss $L(\bm{\theta})$ are minimized and converge in a modest number of training epochs. The TP penalty term in the cost function~\eqref{eq:trace_penalty} helps the LPDO to converge to the correct normalization.}
    \label{fig:training_lpdo_example}
\end{figure}

In this section we investigate the effectiveness of the proposed tensor network noise characterization technique, and analyze how its performance scales with the tomographic dataset size, the number of qubits and the accuracy in estimating noise coefficients. 
Importantly, as discussed in Sec.~\ref{sec:dataSampling}, we stress again that in our analysis we consider realistic scenarios with a limited number of experimental settings and multiple shots per setting, and we find that this is sufficient to provide a reliable approximation of the noisy process. \mar{In all the numerical results reported below, the tomographic data for the channel reconstruction is obtained by sampling, for each qubit, input states from the SIC set of four states defined in Eq.~\eqref{eqn:SICstates}, and measurements from the Pauli basis as described in Sec.~\ref{sec:dataSampling}.} 

From the classical computational viewpoint, the number of trainable parameters in the LPDO scales as $\order{n\chi_{\kappa} \chi_{b}^2}$, so the learning process remains efficient as long as the bond dimensions are small, which is the case for our task of characterizing shallow noisy operations (the largest Kraus and virtual bond dimensions used in our simulations are, respectively, $\chi_\kappa = 16$ and $\chi_b = 4$). Indeed, our largest training experiment with an LPDO of $n=20$ qubits, with bond dimensions $\chi_b =\chi_\kappa=4$, on a tomographic dataset consisting of $N=10^6$ samples, can be run in about one hour on a laptop. 

In Fig.~\ref{fig:training_lpdo_example}, we report an example of the characterization of a brickwork depolarizing noise channel for $n=10$ qubits with $N=10^6$ shots, using an LPDO with $\chi_{\kappa}=2$, $\chi_b = 16$. The reconstruction error~\eqref{eq:characterisation_error} and the loss function evaluated on a test set of samples both decrease along the training process and converge to a minimum value in a few training epochs. Also, after starting from the correct value (see Sec.~\ref{ssec:lpdo_initializatio}), the TP penalty~\eqref{eq:trace_penalty} in the loss function enforces the LPDO to converge to the correct normalization $\Tr[\Lambda_{\bm{\theta}}] \approx 2^n$ by the end of training.

In all the analyses presented below, we show the results obtained with the trained LPDOs $\Lambda_{\bm{\theta}_{\text{opt}}}$ attaining the lowest test error~\eqref{eq:loss_function} during the training process, a measure which is accessible in real experiments and does not require knowledge of the process under characterization. In all experiments with brickwork depolarizing channels, $\chi_b=2$ and $\chi_\kappa=16$ were used, while $\chi_b=4$ and $\chi_\kappa=4$ were used for experiments involving sparse Pauli-Lindblad noise.

\subsection{Accuracy vs. number of shots}
\begin{figure}[t]
    \centering
    \includegraphics[width=0.475\textwidth]{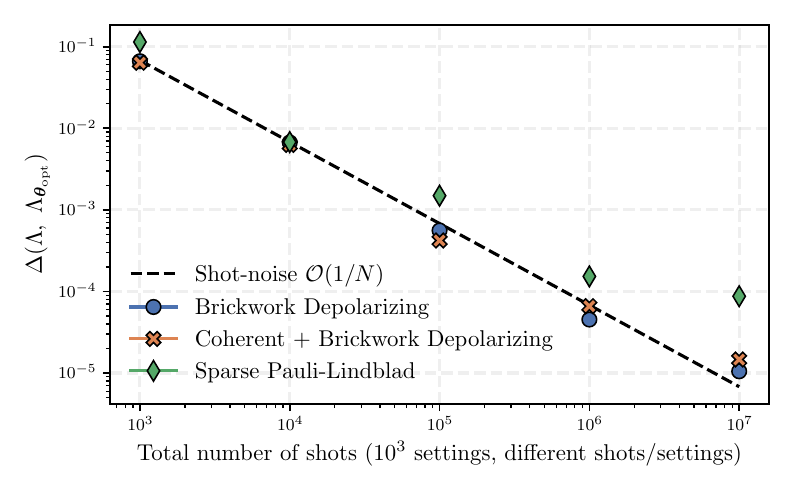}
    \caption{Frobenius distance between true and reconstructed noise channel as a function of the number of shots, for the different noise models introduced in Sec.~\ref{sec:tomography_experiments}, and with $n=10$ qubits. We also plot the expected ``shot-noise'' scaling, decreasing as $1/N$. \mar{Each point in the plot shows the best value obtained in three different training runs of the LPDO initialized with different parameters, but on the same training dataset. We note that all training runs eventually converge to similar performances.}}
    \label{fig:acc_vs_shots}
\end{figure}

In order to evaluate the viability of tensor network noise learning with random tomographic settings in realistic experimental scenarios, we start by analyzing how the reconstruction accuracy behaves with the size $N$ of the tomographic dataset. In particular, we consider a fixed budget of $N_{\text{set}} = 10^3$ experimental settings, and vary the number of shots allocated to each measurement setting, $N_\text{shots} \in \{1, 10, 10^2, 10^3, 10^4\}$. In Fig.~\ref{fig:acc_vs_shots}, we report results for the characterization procedure of the three realistic noise models discussed in Sec.~\ref{sec:tomography_experiments}, for a system of $n=10$ qubits. 

Interestingly, the reconstruction accuracy follows a shot-noise behavior ---we consider the square of the usual shot-noise scaling $\sqrt{N}$ to account for the square in the definition of the reconstruction error $\Delta$~\eqref{eq:characterisation_error}---, which signals that the learning procedure is able to take full advantage of additional tomographic samples. However, when the number of shots per setting is large enough $N_\text{shots} = 10^4$ ($N=10^7$), the reconstruction accuracy starts deviating from the shot-noise scaling, at which point it would be beneficial to increase the number of settings rather than the shots per setting. 

This is especially evident for the sparse Pauli-Lindblad noise model, for which not only the training yields in general a slightly lower reconstruction accuracy, but the Frobenius distance also displays a significant deviation at large number of shots. We believe such behavior to be a consequence both of the intrinsically more complex structure of the Sparse Pauli-Lindblad noise, and also of this channel being more noisy overall (see the noise coefficients in Fig.~\ref{fig:spl_coeffs} and Fig.~\ref{fig:mpo_noise__coeffs}). In Appendix~\ref{app:strong_depol_results} we report results for the characterization of brickwork depolarizing noise with stronger intensity $p=0.1$ (much larger than current two-qubit error rates~\cite{KimIBMUtility2023, mckay2023benchmarking}). The analysis is in agreement with similar but simpler results in~\cite{Torlai2023}, where the reconstruction accuracy was found to decrease in the presence of stronger noise sources. This could be understood as a consequence of the decrease of visibility of the useful signal with increasing noise, indicating that either more resources or a more fine-tuned training routine are needed to distinguish the signal from a background white noise. 

\subsection{Accuracy vs. number of qubits}
\begin{figure}[t]
    \centering
    \includegraphics[width=0.475\textwidth]{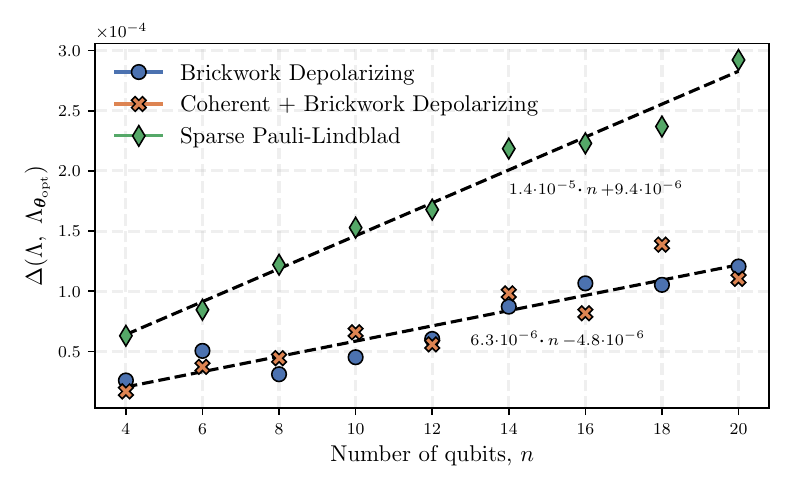}
    \caption{Frobenius distance between true and reconstructed noise channels as a function of the number of qubits, for the different noise models introduced in Sec.~\ref{sec:tomography_experiments}, and with $N_\text{set}=N_\text{shots}=10^3$ (then, $N=10^6$). The dashed lines are linear fits with parameters reported in the figure. \mar{Each point in the plot shows the best value obtained in three different training runs of the LPDO initialized with different parameters, but on the same training dataset. We note that all training runs eventually converge to similar performances.}}
    \label{fig:acc_vs_qubits}
\end{figure}
We now turn our attention to the investigation of the behavior of the reconstruction accuracy as a function of size of the system.

In Fig.~\ref{fig:acc_vs_qubits}, we report the accuracy obtained with $N=10^6$ shots on systems of varying size, up to $n=20$ qubits. We observe a favorable linear scaling of the reconstruction error with the number of qubits $n$ for all noise models considered, which indicates the feasibility of the proposed approach for characterization purposes on near-term devices with a limited amount of qubits. Overall, the results in Fig.~\ref{fig:acc_vs_shots} and Fig.~\ref{fig:acc_vs_qubits} suggest the use of linearly larger tomographic datasets to compensate for the linear decrease in reconstruction accuracy for larger system sizes.

\mar{\subsection{Accuracy in the presence of SPAM errors}
\label{ssec:spam-test}
In this section we show how the proposed noise characterization method can be used also in the presence of SPAM errors, by combining it with techniques aimed at characterizing such state preparation and measurement noise. In particular, we employ the quantum detector tomography (QDT) procedure described in~\cite{Cattaneo2023} to first reconstruct the noisy POVM effects that are actually implemented on the device, and then use such reconstructed effects in the noise characterization procedure. In fact, if state preparation errors are small compared to the other sources of error ---as it is usually the case in current quantum hardware, we observe that the use of measurement tomography alone is already sufficient to recover the reconstruction accuracy obtained in the SPAM-free regime. 

Quantum detector tomography is implemented by executing a set of circuits implementing only state preparation and measurement instructions. Assuming state preparation errors to be negligible compared to measurement errors, by probing the chosen POVM with a set of informationally-complete states, one can realize a tomography of the quantum detector and hence reconstruct the real noisy effects $\smash{\Pi_{\zeta}(\beta) \rightarrow \tilde{\Pi}_{\zeta}(\beta)}$ composing the POVM. These effects are then used in the loss function~\eqref{eq:loss_function} to drive the noise characterization process. In the experiments below, QDT is run using a set of $4$ informationally-complete state to reconstruct the 6-outcome POVM obtained by performing Pauli measurements. This requires a total of $4 \times 3 = 12$ circuits, each of which is executed with $10^4$ shots. Note that the reconstruction of the effects is itself only approximate, with better performance obtained with larger measurement budgets~\cite{Cattaneo2023}. Additionally, the QDT procedure is run assuming an ideal preparation of the input states, while these are in fact also subject to errors. These two effects combined, namely the limited measurement budget and the occurrence of unknown preparation errors, then result in an imperfect reconstruction accuracy of the POVM effects. 

In Fig.~\ref{fig:acc_vs_spam} we report the simulation results obtained by characterizing the sparse Pauli-Lindblad noise on a system of $n=10$ qubits in the presence of realistic SPAM noise, with and without employing QDT to mitigate measurement errors. For the sake of simplicity we consider only incoherent errors: both state preparation and measurement errors are parameterized as single-qubit depolarizing channels, with state preparation having depolarizing strength $\smash{p_\text{prep} = 10^{-4}}$ (which is a reasonable value for single-qubit gates on near-term computers), and measurement error having $\smash{p_\text{meas} \in \{10^{-4}, 10^{-3}, 10^{-2}, 10^{-1}\}}$. 
\begin{figure}[t]
    \centering
    \includegraphics[width=\linewidth]{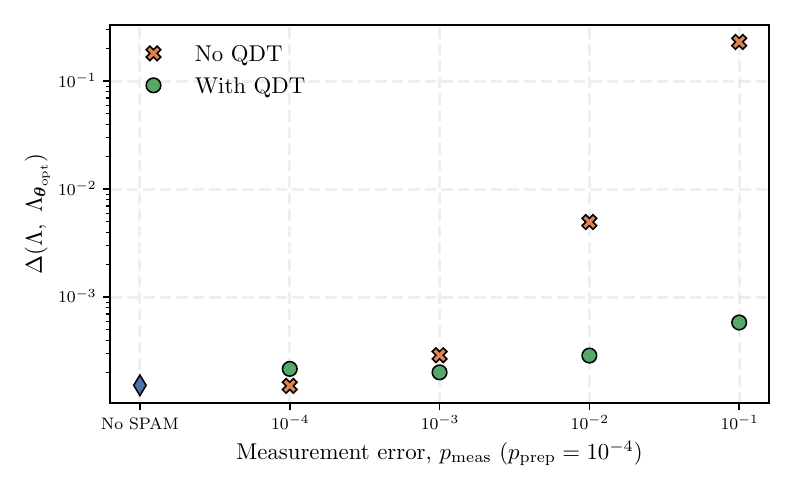}
    \caption{Effect of SPAM errors on the noise characterization procedure on $n=10$ qubits subject to sparse Pauli-Lindblad noise, with $N_\text{set}=N_\text{shots}=10^3$. Both state preparation and measurement errors are parameterized as single-qubit depolarizing channels with intensities $p_\text{prep}$ and $p_\text{meas}$, respectively. Measurements error are mitigated by using quantum detector tomography (QDT)~\cite{Cattaneo2023} with $10^4$ shots to reconstruct the noisy POVM effects.}
    \label{fig:acc_vs_spam}
\end{figure}

When measurement errors are large, the noise learning procedure is unable to provide an accurate description of the noisy evolution, but this can be readily solved by using QDT to calibrate the device and train the LPDO using the reconstructed noisy effects. When SPAM errors are small enough instead, noise characterization obtains good reconstruction accuracy irrespective of the use of QDT. This can be understood by noticing that the true and noisy effects are now very close to each other, and QDT is unable of precisely distinguishing them using a limited number of shots. Additionally, in the regime where state preparation and measurement errors are of the same order of magnitude, QDT yields incorrect noisy effects since it was run assuming ideal state preparations, which can then impact the noise learning procedure. This issue may be solved by using self-consistent characterization protocols~\cite{Cattaneo2023, Nielsen2021}, but we leave this as a subject of future studies.

Overall, our results not only indicate the proposed tensor-network noise learning procedure is stable against small SPAM errors, but also that is can be straightforwardly combined with existing detector tomography methods to calibrate the measurement apparatus and cancel the effects of large measurement errors.}

\subsection{Reconstructed noise coefficients}
\begin{figure}[ht]
    \centering
    \includegraphics[width=0.475\textwidth]{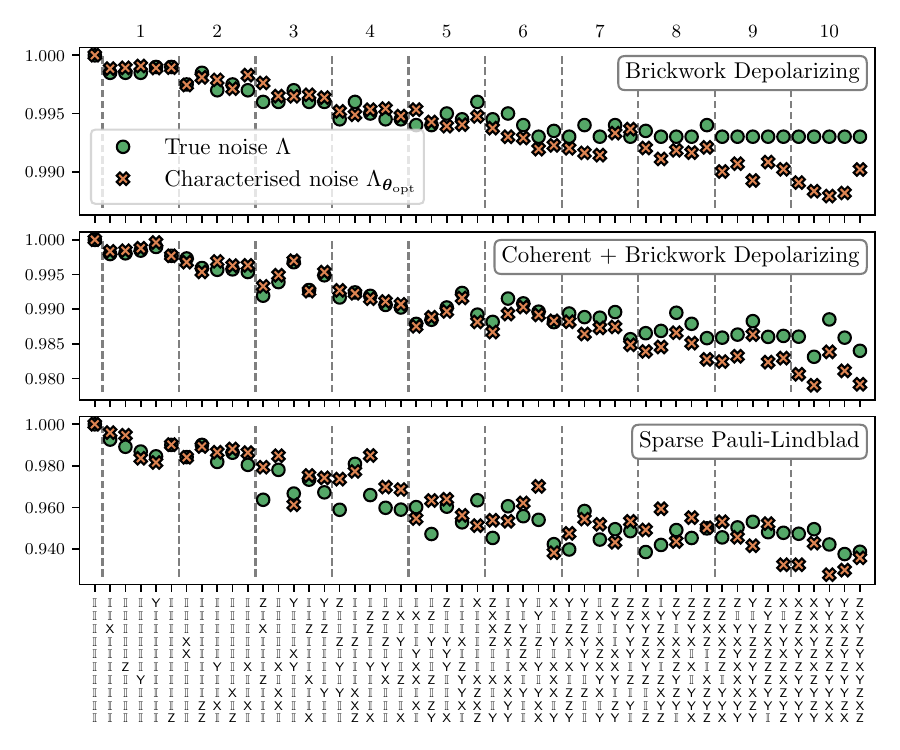}
    \caption{Some diagonal coefficients of the MPOs in Pauli transfer matrix~\eqref{eq:ptm_coeffs} of the true (green circles) and reconstructed (orange crosses) noise channels, for the different noise models introduced in Sec.~\ref{sec:tomography_experiments} and $n=10$ qubits. The numbers above the plot indicate the Pauli weight (i.e. the number of non-identities) of the operators. The mismatch between the green circles and orange crosses is due to the error in channel tomography, obtained with a training set of $N=10^{7}$ ($N_{\text{set}}=10^3$, $N_\text{shots}=10^4$) samples.}
    \label{fig:mpo_noise__coeffs}
\end{figure}

While the Frobenius distance captures the overall difference between the two LPDOs, we investigate more physical figures of merit as well, such as specific coefficients within the tensor network representing the noise channel. 

In particular, let us transform the LPDO representing the channel into an MPO, i.e., let us switch to the superoperator representation of the channel (see Appendix~\ref{app:LPDO_to_MPO} for details). Moreover, we perform a suitable change of basis such that this MPO is written in the basis of Pauli matrices, and then consider the coefficients in the Pauli transfer matrix representation of the noise channel
\begin{equation}
    \label{eq:ptm_coeffs}
    c_{ij} = \frac{1}{2^n}\Tr\qty[P_i~\mathcal{N}\qty[P_j]]\,,
\end{equation}
where $P_{i,j}$ are $n$-qubit Pauli operators.

In Fig.~\ref{fig:mpo_noise__coeffs}, we report some values of these coefficients for both the true and reconstructed noise channels, for the different noise models introduced in Sec.~\ref{sec:tomography_experiments} defined on $n=10$ qubits. As it is clearly unfeasible to investigate all the $4^{2n}$ Pauli coefficients $c_{ij}$, we restrict our analysis to diagonal terms ($i=j$), and report data for some randomly sampled Pauli strings having different Pauli weight (number of non-identities). Note that while for the incoherent brickwork depolarizing~\eqref{eq:totalDepChannel} and sparse Pauli-Lindblad noise models~\eqref{eqn:PauliLindbladNoise} the Pauli Transfer matrix is indeed diagonal, this is not the case for coherent brickwork depolarizing channel which also has non-diagonal elements.

First of all, we note that all the reconstructed noise channels display the correct necessary behavior for trace preservation, as the coefficient belonging to the zero-weight Pauli string (all identities) is correctly normalized to one, as it holds $\Tr[\mathbb{I}~\mathcal{N}[\mathbb{I}]]/ 2^n = 1$.
More importantly, we observe that the errors in the learned noise coefficients are relatively small, with a typical error of order $10^{-3}$ in all cases analyzed.. Interestingly, we note that larger coefficients belonging to low-weight Pauli strings and larger noise levels are easier to learn, a fact which we will investigate more deeply in future studies. We have also checked that the accuracy of the method still holds when characterizing stronger noise, as discussed in detail in Appendix~\ref{app:strong_depol_results}. 

Overall, these results provide an additional and more direct evidence of the potential of the tensor-network approach to characterize noisy processes. 

\section{Application to error mitigation \label{sec:errorMit}}
Despite their broad applicability and straightforward definition, distance measures like the norm in Eq.~\eqref{eq:characterisation_error} may not be directly relevant for practical scenarios when one is interested in applying error mitigation techniques using characterized noise. For example, when it comes to calculating rigorous bounds for, e.g., estimation errors in experiments, the use of distances may lead to very loose bounds of little practical use~\cite{Filippov2023}.

In this section, we test the proposed noise learning procedure on the very timely task of error mitigation, showing how the proposed approach is able to provide accurate enough descriptions of the noise processes to achieve good noise-free estimates of expectation values when used in tandem with error mitigation techniques.

\subsection{Tensor network error mitigation strategy}
\label{sec:mitigationStrategy}
The error mitigation strategy we adopt in this work is the tensor-network error mitigation (TEM) algorithm recently introduced by some of the authors~\cite{Filippov2023}. TEM relies on the (ideally perfect) characterization of the noise channels that affect the quantum circuit. This characterization is then employed to invert and cancel the effect of the noise channels, in the same spirit as in one of the most successful methods for quantum error mitigation, probabilistic error cancellation (PEC)~\cite{Temme2017,Endo2018,vanDerBerg2023}. At variance with PEC, TEM is applied completely in post-processing and, moreover, it provides a quadratic advantage in the sampling overhead with respect to the former~\cite{Filippov2023}. It also provides a sampling advantage with respect to Zero-Noise Extrapolation with Probabilistic Error Amplification (ZNE-PEA). In fact, for specific cases, it can be shown that its sampling overhead is optimal~\cite{Filippov2024}.

Suppose that the circuit we want to run on the quantum computer is composed of $M$ layers represented by the ideal unitaries 
\begin{equation}
\label{eqn:ideal_sim}
    \mathcal{C}_\text{ideal}=\mathcal{U}_M\circ \ldots \circ \mathcal{U}_1\,.
\end{equation} 
However, due to inevitable noise in the quantum processor, the evolution we implement on hardware is instead given by 
\begin{equation}
\label{eqn:noisy_sim}\mathcal{C}_\text{noisy} = \mathcal{N}_M\circ\mathcal{U}_M\circ\,\ldots\,\circ\mathcal{N}_1\circ\mathcal{U}_1\,,
\end{equation} 
where $\mathcal{N}_j$ is the noise channel associated with the ideal unitary operation $\mathcal{U}_{j}$ in the $j$-th layer. After running the noisy circuit on hardware and obtaining the final outcome through a proper measurement procedure, our goal is to improve the accuracy of the outcome by mitigating the detrimental effect of the noise channels $\mathcal{N}_j$. The way we can achieve this through TEM is the following.  

First, we characterize the noise channels $\mathcal{N}_j$ in tensor network formalism. This characterization should be as accurate as possible and, crucially, we should be able to characterize the same layers we are using during the actual execution of the quantum circuit in Eq.~\eqref{eqn:noisy_sim}. That is, the noise on the hardware should not change in the time between characterization and execution. Then, by computing the inverse of the noise channels $\mathcal{N}_j^{-1}$ \mar{(see Appendix~\ref{app:mpo_inversion} for more details)}, we can finally post-process the informationally complete measurement results obtained from the noisy state by applying the non-physical map
\begin{equation}
\label{eqn:tem_map}
\mathcal{C}_{\text{TEM}} = \mathcal{U}_M\circ \ldots\circ\,\mathcal{U}_1\circ\,
\mathcal{U}_1^{-1}\circ\,\mathcal{N}_{1}^{-1}\circ\ldots\circ\,\mathcal{U}_M^{-1}\circ\,\mathcal{N}_{M}^{-1}\,,
\end{equation}
for which it is easy to see that $\mathcal{C}_\text{TEM} \circ\, \mathcal{C}_\text{noisy} = \mathcal{C}_\text{ideal}$, that is, we recover the ideal output. 

The mitigation map $\mathcal{C}_\text{TEM}$ is represented as a tensor network and it is thus computed, i.e. contracted, on a classical computer. If $\mathcal{C}_\text{TEM}$ were as complex ---from a tensor network perspective--- as $\mathcal{C}_\text{ideal}$, then TEM would not be of any use, since we would only be able to mitigate noise through classical tensor network methods if we were also able to directly compute the evolution driven by $\mathcal{C}_\text{ideal}$ through the same techniques. The core idea of TEM, however, is that only the inverse of the aggregated noise in the circuit must be classically simulated. If the noise in the channels $\mathcal{N}_j$ is small enough, then the post-processing map approaches the identity operator $\mathcal{C}_\text{TEM} \approx \mathcal{I}$, and thus its contraction can be computed efficiently through tensor network methods, even if we are dealing with a large number of qubits. We refer the interested readers to the original paper~\cite{Filippov2023} for more details and discussions about TEM. 

\subsection{Numerical results}
\label{sec:mitRes}
\begin{figure*}[!t]
    \centering
    \includegraphics[width=\textwidth]{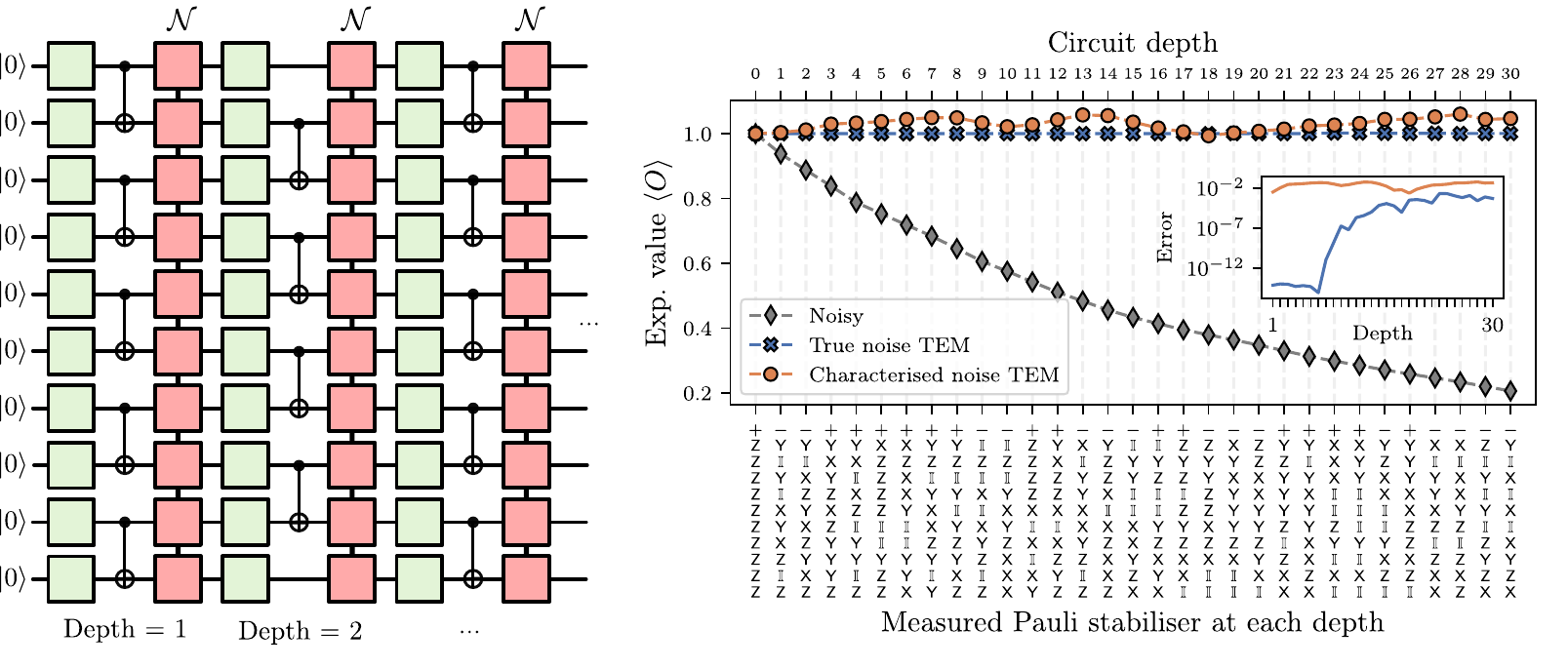}
    \caption{Left: Schematics of the noisy Clifford circuits we consider into account. One ideal layer consists of a random single-qubit Clifford operations (green squares) followed by a (even or odd) CNOT layer. Each ideal layer is followed by the noise channel $\mathcal{N}$, which is a sparse Pauli-Lindblad noise channel introduced in Sec.~\ref{sec:pauliLindbladNoiseModel}, with coefficients sampled in order to resemble real experiments on IBM computers, see Appendix~\ref{sec:appendixSparsePauliLindblad} for details. Right: Results of the numerical experiment of error mitigation applied to the noisy circuit on the left. The expectation value of a different Pauli stabiliser at each circuit depth is shown, for either the unmitigated noisy circuit (gray diamonds), the mitigated circuit through TEM based on the true noise channel (blue crosses), and the mitigated circuit through TEM based on the reconstructed noise channel (orange circles) using a training set of $10^6$ samples. Inset: mismatch between the mitigated results and the true result (equal to 1 for all depths), for either TEM based on the true noise channel (blue line) or the TEM based on the reconstructed noise channel (orange line). The blue line is different from zero due to the bond dimension truncation in the TEM method only. The orange line, in contrast, comprehends errors arising from the inaccuracy in the channel reconstruction (dominant contribution), the bond dimension truncation, and also errors in the inversion procedure of the MPO representing the noise necessary to run TEM. Nonetheless, the characterization procedure is able to provide a remarkably accurate description of the noise, so that TEM is able to provide almost ideal noise-mitigated values even at large depth.}
    \label{fig:tem-example}
\end{figure*}

To test the noise characterization method, we numerically simulate a noise mitigation experiment on $n=10$ qubits in which we employ the noise channel returned by the characterization protocol together with TEM to mitigate the noisy circuit depicted in Fig.~\ref{fig:tem-example} (left). 

The circuits we analyze consist in a repeated structure of operations sampled from Clifford gates, which allows for an easy computation of ideal noise-free expectation values from the circuit~\cite{nielsen_chuang_2010}. In order to study the accumulation of errors in deep circuits due to noise happening on several computational layers, we run the tensor mitigation strategy on several circuits of different depths, obtained by iteratively appending additional layers one after another.

One step of the ideal (noise-free) circuit comprises a layer of random single-qubit Clifford gates followed by a layer of CNOTs, with the CNOT gates in each layer acting either on even or odd links between the qubits, depending on the step. Note that such alternating brickwork circuit structure is of practical interest as it can be used, for example, to study properties of many-body quantum systems via Trotterized evolution, see e.g.~\cite{KimIBMUtility2023,Keenan2023}. At the end of the circuit, we assume we are measuring the stabilizer Pauli operator $O$ having expectation value $\expval{O} = +1$, which can be calculated by evolving the initial Pauli string $Z^{\otimes n}$, whose $+1$ eigenstate is the initial ground state $\ket{0}^{\otimes n}$ of the computation, with the Clifford operations in the circuit. 

To take noise into account, we assume that each ideal circuit layer is followed by a noise channel $\mathcal{N}$, the effects of which we aim to mitigate through TEM in post-processing. For these experiments, we set the noise channel $\mathcal{N}$ to be a sparse Pauli-Lindblad noise~\eqref{eqn:PauliLindbladNoise} with the coefficients as in Appendix~\ref{sec:appendixSparsePauliLindblad}, sampled to resemble publicly available data by IBM on recent experiments leveraging SPL noise models~\cite{KimIBMUtility2023, vanDerBerg2023}. Notably, as discussed in Sec.~\ref{sec:pauliLindbladNoiseModel}, since such noise is also a Clifford map, its effects on the output of the circuit can be computed efficiently.

We perform the TEM experiments with both the exact noise model used in the noisy simulations, and with the noise model obtained with the characterization procedure using a total of $10^6$ random measurement shots, as in Fig.~\ref{fig:acc_vs_shots}. In order to use the characterized LPDO of the noise with TEM~\eqref{eqn:tem_map}, we first transform it into an MPO and then compute its inverse by combining the explicit linear-algebra based approach proposed in~\cite{Guo2022} together with an additional variational minimization. We refer to Appendix~\ref{app:LPDO_to_MPO} and~\ref{app:mpo_inversion} for further details on the LPDO to MPO transformation and inversion of an MPO, respectively. In the simulations below the bond dimension of the MPO used to represent the tensor error mitigation map~\eqref{eqn:tem_map} is $\chi = 200$.

As discussed before, the mean values of the different Pauli operators $O$ considered in the right panel of Fig.~\ref{fig:tem-example} are always equal to $+1$ for the ideal noise-free circuits, as for each step we are measuring the Pauli operator stabilized by that circuit. The same expectation values but for the noisy circuit are also shown (grey diamonds) up to 30 steps, with the signal almost vanishing at the last step.

In the right panel of Fig.~\ref{fig:tem-example}, we show the TEM-mitigated results of the mean values of the Pauli operators using as an input for TEM either the true noise channel (blue crosses), which we can perfectly know only in a numerical experiment, or the reconstructed noise channel obtained through tensor-network-based noise characterization (orange circles). The difference between the blue crosses and the ideal value $+1$ is due to the truncation of the bond dimension in the TEM method and, as shown in the inset of Fig.~\ref{fig:tem-example} (right), it is small and noticeable only for higher circuit depths. In other words, for all practical purposes, TEM reproduces the exact ideal result in the first circuit steps. In contrast, there is a visible difference between the ideal noise-free estimates and the mitigated ones obtained with the characterized noise model. However, in the inset, we observe that such mismatch is always of the order of  $10^{-2}$ and, importantly, it does not increase with the circuit depth, so we are able to recover an almost perfect result even at step 30, where the noise almost wiped out the signal entirely. This is remarkable, given that three different sources of error are at play at the same time: (\textit{i}) reconstruction error inherent to the noise learning procedure, (\textit{ii}) errors in the inversion of the MPO of the characterized noise, and eventually (\textit{iii}) truncation errors introduced by TEM to compute the tensor network mitigation map~\eqref{eqn:tem_map}, with the first one dominating over the other two.

Our results thus show that the tensor-network-based noise characterization scheme studied in this work can provide an accurate description of the noise even with a modest number of training experimental data, with direct applications in error mitigation techniques that rely on the knowledge of the noise. 

\section{Conclusions}
\label{sec:conclusions}
Accurate noise characterization is of utmost importance for attaining the best performance out of near-term quantum computers, and especially for state-of-the-art error mitigation methods, many of which rely on accurate knowledge of the noisy gates physically applied~\cite{Filippov2023, vanDerBerg2023}. Standard process tomography~\cite{nielsen_chuang_2010}, however, is unfeasible in the era of quantum utility, as circuit layers with tens or hundreds of qubits would require a huge amount of tomographic resources (e.g., experimental setups, state preparations, measurement shots, etc.), the scaling of which is exponential in the number of qubits. 

In this work, we propose a protocol for noise characterization based on the tensor network procedure introduced by Torlai \textit{et al.}~\cite{Torlai2023}. Our method not only avoids the exponential scaling of measurement resources by sampling the different possible tomographic settings in a randomized way, but it also enables an efficient, meaningful, and scalable description of the reconstructed noise channel by means of tensor network techniques (more specifically, a locally-purified density operator structure, LPDO) with low bond dimension. The investigated method does not require any twirling of the noise maps~\cite{Wallman2016}, and is therefore suited to learn generic noisy processes. As the output of our protocol is a tensor network representation of the noise channel, it can be directly used as an input for the tensor network error mitigation (TEM) algorithm recently introduced in~\cite{Filippov2023}. 

\mar{Whereas the original proposal in~\cite{Torlai2023} mainly focused on learning unitary processes coming from arbitrarily deep quantum circuits, the originality of our approach lies in specializing the channel tomography technique to the case of learning shallow noise maps that accompany imperfect circuit layer instructions. This makes the procedure practical, as it requires low bond dimensions, and highly relevant for many noise-aware mitigation protocols. Additionally, we extensively tested the method in several scenarios that are experimentally relevant, including the effect of SPAM errors. We also compared our method with a similar proposal based on a local tomographic strategy, which demonstrated worse performance. Finally, we tested the protocol for error mitigation, which is crucial for the success of near-term quantum computation and the primary reason for needing need noise characterization}

Our protocol was tested through several numerical experiments for realistic multi-qubit correlated noise model learning. We specifically addressed three different noise channels that are of great relevance for current quantum computation, namely the sparse Pauli-Lindblad noise model~\cite{vanDerBerg2023}, the incoherent depolarizing noise model with crosstalk, and the depolarizing noise model with crosstalk and coherent errors. 

To assess the accuracy of the reconstruction, we used two figures of merit: the Frobenius distance between the true and reconstructed LPDOs, and the difference between the elements of the true and reconstructed superoperators expressed in the Pauli basis. We found that a limited and experimentally feasible number of shots (around $10^6$ per characterization experiment) suffices for accurate noise channel characterization. We also explored how accuracy scales with the number of shots and qubits, observing favorable linear behavior in both cases.

\mar{Importantly, we have also tested the efficacy of the method in the presence of SPAM errors, demonstrating its resilience against small errors and how it can be combined with existing detector tomography techniques to mitigate undesired measurement errors and retain good channel reconstruction accuracies.}

Moreover, we benchmarked the method with the timely and relevant task of quantum error mitigation. Specifically, we used the output of the noise characterization procedure as input for the TEM protocol to mitigate measurement outcomes from a noisy quantum circuit. We showed that TEM with characterized noise is able to provide mitigated expectation values with good accuracy (relative error of the order of $10^{-2}$), even on deep circuit instances with tens of layers.


Summarizing, our analysis suggests that the tensor network noise characterization protocol may be an valuable tool for error mitigation for near-term quantum computers. The accuracy of our method is corroborated by the precision with which we can both reconstruct the noise channel as a tensor network and recover the ideal result of a noisy circuit when we employ this channel in conjunction with TEM.  

\begin{acknowledgments}
The authors would like to thank Chris J. Wood for interesting discussions about noise characterization on near-term quantum computers. The code for the noise characterization method based on tensor networks presented in this work is integrated into \emph{Aurora}, a proprietary quantum chemistry platform developed by Algorithmiq Ltd.
\end{acknowledgments}

\section*{Author contributions}
GGP proposed the idea and supervised the project together with DC. SM and MC conducted the research. SM wrote the main numerical code for this work. SM and MC ran numerical simulations of this code on classical computers and wrote the first draft of the manuscript. SF, MR provided the code and support for TEM mitigation, SF provided support for TN parametrization. All authors
contributed to scientific discussions and to the writing of the manuscript.

\appendix

\section{Tensor network details}
\label{sec:tensorNetDetails}

\subsection{Explicit expression of the quantum channel in tensor network notation}
\label{sec:LPDO_definition}
The diagrammatic representation of the LPDO in Fig.~\ref{fig:TNstructure} can be written in tensor network notation with explicit indices as
\begin{equation}
\label{eqn:LPDO_definition}
\begin{split}
&[\Lambda_\mathcal{N}]_{\boldsymbol{a},\boldsymbol{a'}}^{\boldsymbol{b},\boldsymbol{b'}} = \sum_{\mu_1,\ldots,\mu_{n-1}}\sum_{\nu_1,\ldots,\nu_{n-1}}\sum_{\kappa_1,\ldots  ,\kappa_n} \prod_{j=2}^{n-1} \\  & \quad[A_1]_{b_1 a_1}^{\mu_{1}\kappa_1}[A_1^\dagger]_{a'_1 b'_1}^{\nu_{1}\kappa_1}\,\ldots\,[A_j]_{b_j a_j}^{\mu_{j-1}\mu_j\kappa_j}[A^\dagger_j]_{a_j'b_j'}^{\nu_{j-1}\nu_j\kappa_j}\\
& \quad \ldots\,[A_n]_{b_n a_n}^{\mu_{n-1}\kappa_n}[A_n^\dagger]_{a_n'b_n'}^{\nu_{n-1}\kappa_n}\,.
\end{split}
\end{equation} 
\mar{The coefficients $\smash{[A_j]_{b_j a_j}^{\mu_{j-1}\mu_j\kappa_j}}$ are obtained by choosing a basis for the single-qubit operators, and then decomposing each global (i.e., acting on all the qubits) Kraus operator $K_{\kappa}$ into a linear combination of tensor products of single-qubit operators, as given by Eq.~\eqref{eqn:KrausDecTN}. One general approach to split each Kraus operator into single-qubit operators (and the corresponding index $\kappa$ into local indexes $\kappa_1,\ldots,\kappa_n$) is based on a recursive application of the singular value decomposition, and the choice of $\smash{[A_j]_{b_j a_j}^{\mu_{j-1}\mu_j\kappa_j}}$ depends on the specific decomposition one applies, see e.g. Ref.~\cite{SchollwckDMRG2011}. 

In practical cases however, the local decomposition is evident from the structure of the channel under investigation. For example, a general two-qubit Pauli channel reads
\begin{equation}
\begin{aligned}
    \mathcal{P}[\rho] & = \sum_{\kappa=1}^{16} c_\kappa~P_\kappa \rho P_\kappa \\
    & = \sum_{\kappa_1, \kappa_2 =1}^{4} c_{\kappa_1 \kappa_2}~(P_{\kappa_1}\otimes P_{\kappa_2})\,\rho\, (P_{\kappa_1}\otimes P_{\kappa_2})\,,
\end{aligned}
\end{equation}
with $P_{\kappa_i} \in \{\mathbb{I}, X, Y, Z\}$ being single-qubit Pauli matrices. Such channel has Kraus operators $K_{\kappa_1\kappa_2} = \sqrt{c_{\kappa_1 \kappa_2}}\, P_{\kappa_1} \otimes P_{\kappa_2}$ which have a clear local structure. Starting from such decomposition, one can realize that a LPDO representation of the channel as in Eq.~\eqref{eqn:KrausDecTN} is achievable starting from the local tensors
\begin{equation}
\begin{aligned}
    \qty[A^{[1]}]_{\mu\kappa_1} &= 
    \begin{bmatrix} 
    \mathbb{I} & 0 & 0 & 0 \\
    0 & X & 0 & 0 \\
    0 & 0 & Y & 0 \\
    0 & 0 & 0 & Z \\
    \end{bmatrix}\,,
    \\
    \qty[A^{[2]}]_{\mu\kappa_2} &= 
    \begin{bmatrix} 
    \sqrt{c_{11}}\,\mathbb{I} & \sqrt{c_{12}}\,X & \sqrt{c_{13}}\,Y & \sqrt{c_{14}}\,Z \\
    \sqrt{c_{21}}\,\mathbb{I} & \sqrt{c_{22}}\,X & \sqrt{c_{23}}\,Y & \sqrt{c_{24}}\,Z \\
    \sqrt{c_{31}}\,\mathbb{I} & \sqrt{c_{32}}\,X & \sqrt{c_{33}}\,Y & \sqrt{c_{34}}\,Z \\
    \sqrt{c_{41}}\,\mathbb{I} & \sqrt{c_{42}}\,X & \sqrt{c_{43}}\,Y & \sqrt{c_{44}}\,Z \\
    \end{bmatrix}
\end{aligned}\,.
\end{equation}
Note however that such representation is non-unique, due to the inherent gauge freedom of tensor networks (for example, exchanging the two local tensors $A^{[1]} \leftrightarrow A^{[2]}$ give rise to the same channel).

More complex channels arising from combinations of single- and two-qubit channels ---as the ones discussed in the main text--- can be obtained by combining and contracting together the LPDO representation of each of these channels.
}

\subsection{From locally purified density operators (LPDO) to matrix product operators (MPO) \label{app:LPDO_to_MPO}}
\begin{figure*}[ht]
    \centering
    \includegraphics[width=\textwidth]{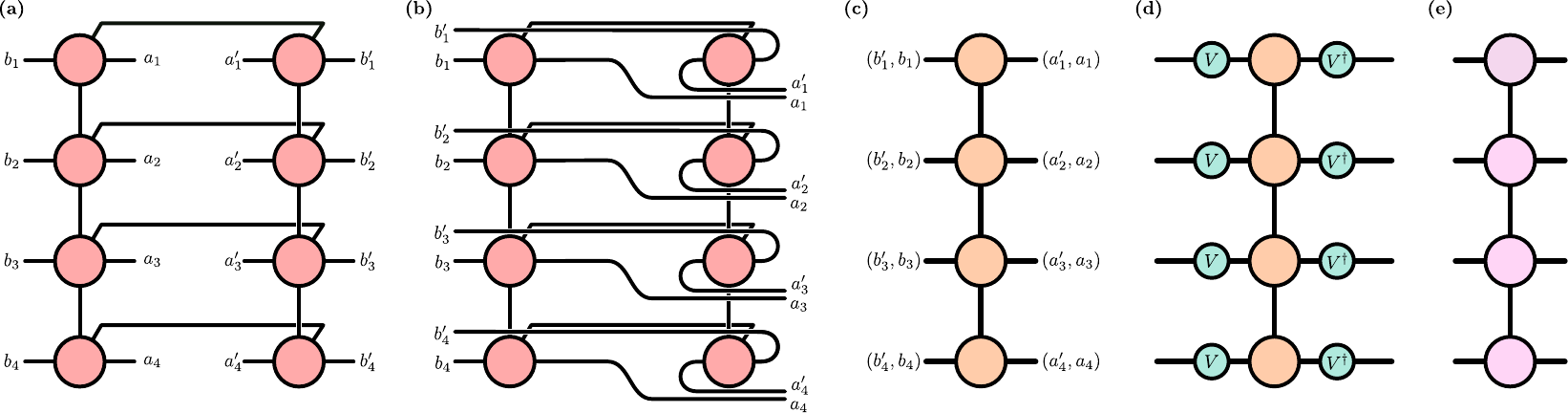}
    \caption{\textbf{(a)} LPDO representation of a quantum channel $\mathcal{N}$, analogously to Fig.~\ref{fig:TNstructure}. \textbf{(b)} indices are reshuffled according to the superoperator representation. \textbf{(c)} The reshuffled indices are merged to give rise to a MPO representing $\mathcal{N}$ as a superoperator. \textbf{(d)} A change of basis transformation is applied locally on each qubit to write the MPO in the Pauli basis (starting from the computational one). \textbf{(e)} Final MPO representation of $\mathcal{N}$ in the Pauli basis.}
    \label{fig:lpdo2mpo}
\end{figure*}

In Fig.~\ref{fig:lpdo2mpo}(a) it is represented the LPDO representation of the quantum channel $\mathcal{N}$ we want to characterize. In many applications (e.g., for running the TEM algorithm) we need the superoperator representation of $\mathcal{N}$ in the Liouville space, which consists of transforming the channel into a matrix acting on the vectorized space of density matrices~\cite{WoodTNGraphical2015}.

In the superoperator formalism, the action of some Kraus operators $K_\kappa$ acting on the state $\rho$ like $K_\kappa \rho K_\kappa^\dagger$ is represented as $K_\kappa \otimes (K_\kappa^\dagger)^T |\rho\rangle\!\rangle$, where $|\rho\rangle\!\rangle$ is a suitable vectorization of the density matrix. Then, the superoperator associated with $\mathcal{N}$ in the tensor network formalism can be easily obtained from the LPDO structure as shown in Fig.~\ref{fig:lpdo2mpo}(b-c): the indices of the Kraus operators acting on the left and on the right of the density matrix are suitably reshuffled and then merged to create a MPO representing $\mathcal{N}$~\cite{WoodTNGraphical2015}. Additionally, for the sake of running the tensor error mitigation (TEM) algorithm described in Sec.~\ref{sec:mitigationStrategy}, we need a MPO superoperator representation of $\mathcal{N}$ in the Pauli basis. To do this we apply on each site of the MPO a local change-of-basis unitary operator that transforms the computational basis into the desired Pauli basis, as depicted in Fig.~\ref{fig:lpdo2mpo}(d). Finally, the MPO we will invert to run TEM is shown in Fig.~\ref{fig:lpdo2mpo}(e).

\section{Local sampling strategies \label{app:local_vs_global}}
\label{sec:localStrategy}
The data sampling employed in this paper is based on the random strategy described in Sec.~\ref{sec:dataSampling}. We have also explored a different strategy that assumes that the correlations between different sites of the tensor network representing the quantum channel $\mathcal{N}$ are only $\ell$-local, thus focusing on the reconstructing of $\ell$-reduced channels only. This strategy is motivated by the similar methods that has been successfully applied to the state tomography of Matrix Product States (MPS) \cite{Cramer2010, Lanyon2017} and mixed states~\cite{Guo2023}.

There is, however, a fundamental difference between the local strategy for state tomography and for process tomography. Suppose that we are employing the Pauli measurements (i.e., 3 different measurement bases per qubit, as discussed in the Sec.~\ref{sec:dataSampling}) for $\ell$-qubit state tomography; then we need $3^\ell$ different experimental tomographic settings, corresponding to 6 different POVM outcomes. As discussed in Sec.~\ref{sec:dataSampling}, for $\ell$-qubit process tomography, in contrast, even in the best possible scenario we need $12^\ell$ settings in order to take into account also the preparation of informationally complete input states. This means that the number of experimental settings grows much faster than for state tomography as a function of the locality $\ell$. For a real experiment on current near-term quantum computers with limited access and capabilities, it is already quite difficult to gather statistics on pretty low locality, for example $\ell=4$ implying $12^4$ different experimental settings (quantum circuits), and absolutely unfeasible to reach locality $\ell=5$. 

Fixing the value of the locality $\ell$, a basic local sampling strategy can be implemented by preparing all the possible tomographic settings on subsets of $\ell$ qubits. Specifically, as described in Sec.~\ref{sec:dataSampling}, if we use a set of $R$ informationally complete input states and Pauli measurements, we will need to execute at least $(3R)^\ell$ different experimental settings. In fact, for a linear chain of qubits, one can see that by using a scheme of correlated preparations and measurements comprising $(3R)^\ell$ settings is enough to provide the necessary $\ell$-local tomographic data on all $\ell$-local reduced channels on neighboring qubits.

For instance, suppose we want to characterize 5 qubits and we choose locality $\ell=3$. Consider experimental settings where the qubits are prepared in a correlated fashion, that is with input states of the form $\rho = \rho_{A} \otimes \rho_{B} \otimes \rho_{C} \otimes \rho_{A} \otimes \rho_{B}$, where $\rho_{A,B,C}$ are sampled from a set of IC states; and also measured on correlated bases, that is with measurement operators of the form  $P = P_{A} \otimes P_{B} \otimes P_{C} \otimes P_{A} \otimes P_{B}$, where $P_{A,B,C}$ are Pauli operators. Using such correlated tomographic scheme, by considering all the $(3R)^3$ tomographic settings obtained by considering all combinations of states $\rho_{A,B,C}$ and measurements $P_{A,B,C}$, one covers the experimental settings needed to reconstruct all the $3$-reduced channels acting on subsets of qubits \{0,1,2\}, \{1,2,3\}, and \{2,3,4\}. This is because if states $\rho_{A} \otimes \rho_{B} \otimes \rho_{C}$ on qubits \{0, 1, 2\} spans all $R^3$ possibilities, then also $\rho_{B} \otimes \rho_{C} \otimes \rho_{A}$ on qubits \{1,2,3\} will span all possibilities, and similarly for $\rho_{C} \otimes \rho_{A} \otimes \rho_{B}$ on qubits \{2,3,4\}. Same goes for the measurement operators. This optimal scheme holds for a linear chain of qubits, for more complex topologies the choice of settings may be different~\cite{Araujo2022}.

Alternatively, we may implement the local strategy by keeping a different locality for the input states and the measurement bases. This is motivated by \textit{lightcone} arguments. That is, if we want to characterize all the $\ell$-local outcomes (i.e., we characterize up to $\ell$-local correlations in the measurements over the $n$ qubits), then the outcomes over $\ell$ qubits will in general depend on the input states over more than $\ell$ qubits, depending on the entangling structure of the channel. For instance, consider a single layer of noisy CNOTs in which crosstalk errors affect only the first neighbors, which is the case treated in the main text and depicted in Fig.~\ref{fig:main-figure}(a). Then, it is easy to see that the outcomes on a single qubit can be influenced by the initial states of at most $4$ qubits. If we aim to characterize $2$-local outcomes instead, these will depend on the input state of at most $6$ qubits. This analysis tells us immediately that the scaling of this lightcone-based local strategy is again quite unfavorable: in the best scenario, we need $4^6 \times 3^2 = 36864$ settings for exactly characterizing $2$-local correlations in the measurement outcomes, which is hardly feasible on current near-term quantum computers. 

Independently of the chosen strategy to collect local tomographic data, one could then still use the same machinery discussed in~\ref{sec:TNoptimization} to train a tensor network for the total channel $\mathcal{N}$ starting however from tomographic data on the $\ell$-local channels. Of course, as a general $n$-qubit quantum channel cannot be written in terms of products of $\ell$-local ones, the reconstruction accuracy of the whole channel will be impacted, with good accuracy reached only when the experiments locality used to collect the tomographic data approaches the actual locality of the channel~\cite{Cramer2010, Lanyon2017}. 

\begin{figure}[ht]
    \centering
    \includegraphics[width=0.475\textwidth]{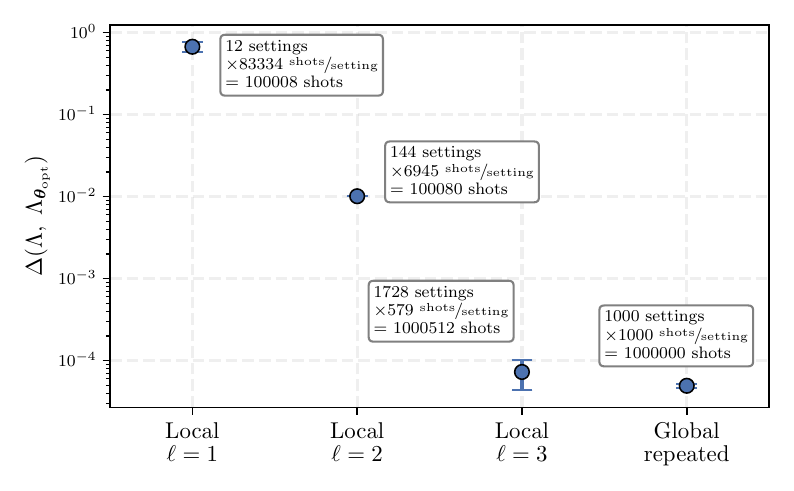}
    \caption{Characterizing the brickwork depolarizing channel with local data. Frobenius distance between true and reconstructed noise channels through the local strategy, for different localities $\ell \in \{1, 2, 3\}$. The number of shots per setting in each scenario is tuned so that all characterization experiments use roughly the same total number of shots ($\approx 10^6$). The results are compared with the global repeated strategy used in the main text (see Sec.~\ref{sec:dataSampling}), consisting of $10^3$ settings and $10^3$ shots per setting, for a total of $10^6$ total shots. The reconstruction error improves by considering larger localities, but the global strategy achieves better performances. Each point in the plot is obtained as the mean value of three different training runs of the LPDO initialized with different parameters, with the error bars being the standard deviation.}
    \label{fig:global_vs_local}
\end{figure}

For completeness, in Fig.~\ref{fig:global_vs_local} we report some numerical results obtained by training the LPDO on local data of different locality $\ell$ obtained using the basic sampling strategy described above, to learn the brickwork depolarizing channel~\ref{sec:incohDep}. As clear from the picture, the accuracy improves when considering a larger locality for the data sampling, but the global strategy still results in better reconstruction performances despite using a smaller number of experimental settings.

\section{Sparse Pauli-Lindblad noise model}
\label{sec:appendixSparsePauliLindblad}
\begin{figure*}[ht]
    \centering
    \includegraphics[width=\textwidth]{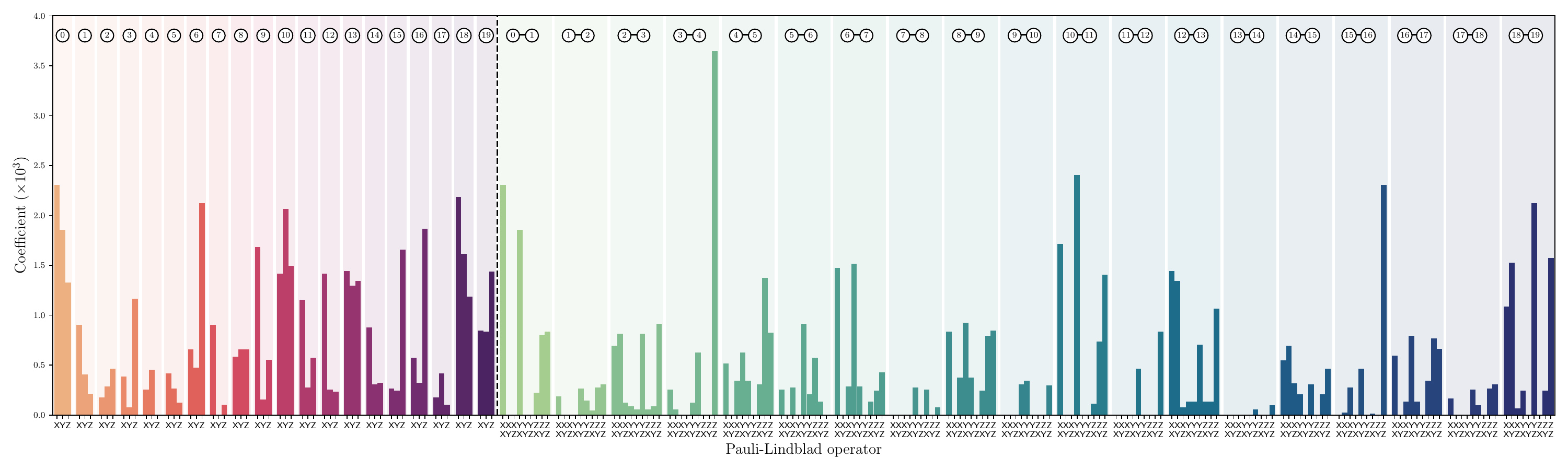}
    \caption{Coefficients of the sparse Pauli-Lindblad noise model~\eqref{eqn:PauliLindbladNoise} considered in the numerical experiments, defined on a maximum of $n=20$ qubits with linear connectivity. We remark that these values are realistic noise coefficients sampled according to publicly available data by IBM on noise characterization run on real superconducting quantum hardware.}
    \label{fig:spl_coeffs}
\end{figure*}

In Fig.~\ref{fig:spl_coeffs} we report the coefficients used in the experiments involving the sparse Pauli-Lindblad noise model, as defined in Eq.~\eqref{eqn:PauliLindbladNoise}. Such coefficients were sampled randomly to match publicly available data by IBM on noise characterization procedures run on superconducting quantum hardware~\cite{KimIBMUtility2023, vanDerBerg2023}. Whenever we consider instances of such noise model on systems with less then 20 qubits ($n<20$), we proceed by simply restricting the noise model to those Pauli-Lindblad operators which act non-trivially on qubits $q \in \{0,\, \ldots,\, n-1\}$.

\section{Strong depolarizing noise\label{app:strong_depol_results}}
In this appendix, we report numerical results for the characterization of a stronger noise channel, namely the brickwork depolarizing channel of Eq.~\eqref{eq:totalDepChannel} but  with noise parameter $p=10^{-1}$, as opposed to $p=10^{-3}$ used in the main text. Note that such error rate is widely larger than those found in already available state-of-the-art quantum computers.

\begin{figure}[ht]
    \centering
    \includegraphics[width=0.475\textwidth]{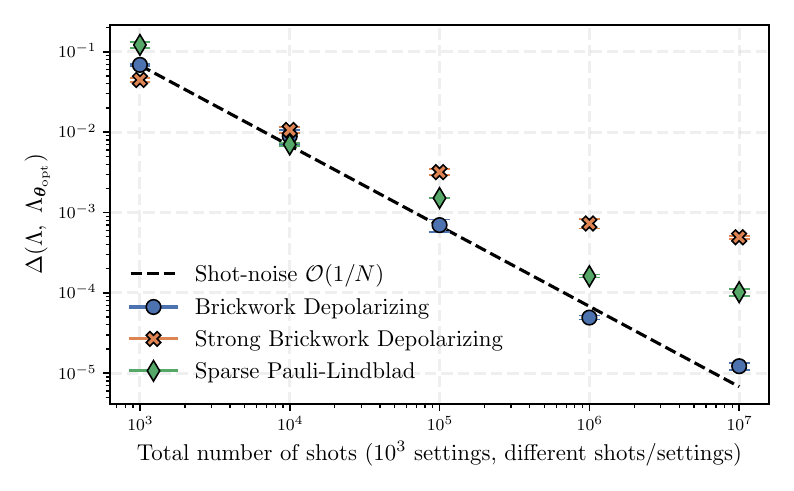}
    \caption{Same as in Fig.~\ref{fig:acc_vs_shots} in the main text, but with the strong incoherent depolarizing noise channel introduced in Sec.~\ref{sec:incohDep} with noise strength $p=0.1$.}
    \label{fig:shots_strong_depol}
\end{figure}

In Figure~\ref{fig:shots_strong_depol} we show the scaling of the Frobenius distance between the true and reconstructed channel as a function of the number of shots, and compare it with the other noise models we have explored in this work. As argued in the main text, we witness a clear dependence of the reconstruction error on the noise intensity, which can be understood as a consequence of the whole learning procedure being unable to distinguish the signal from a background white noise. 

\begin{figure}[ht]
    \centering
    \includegraphics[width=0.475\textwidth]{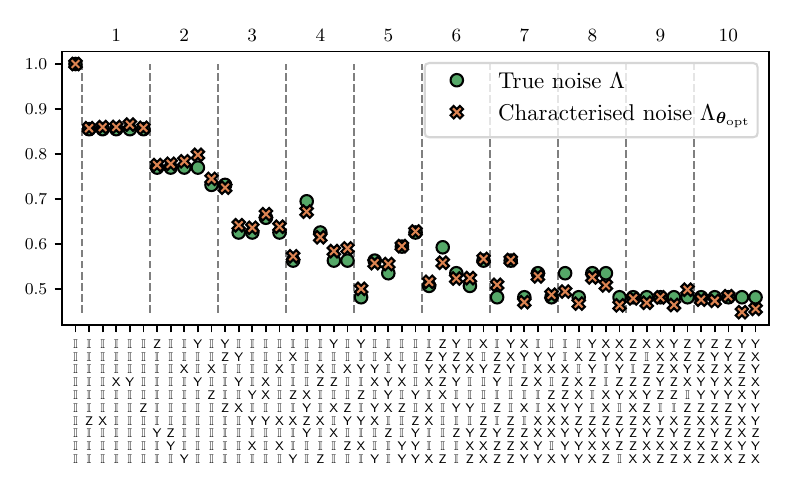}
    \caption{Some coefficients of the MPOs in Pauli transfer matrix of the true (green circles) and reconstructed (orange crosses) noise channels, for the strong incoherent depolarizing noise channel introduced in Sec.~\ref{sec:incohDep}, with $p=0.1$. The numbers above the plot indicate the coefficient of the chosen Pauli operator~\eqref{eq:ptm_coeffs}. The mismatch between the green circles and orange crosses is due to the error in channel tomography, obtained with a training set $N=10^7$ samples. }
    \label{fig:coeffs_strong_depol}
\end{figure}

In addition, in Fig.~\ref{fig:coeffs_strong_depol} we report some coefficients of the true and reconstructed noise channel in the MPO representation and in the Pauli basis. Despite the lesser performance in terms of channel distance, we observe that the accuracy of our reconstruction procedure is however once again remarkable even in the strong noise scenario.

\section{Initialization and optimization of the tensor network}
In this appendix we discuss the custom random initialization of the LPDO tensor network and provide details on the optimization routines to train it.

\subsection{Initialization of the LPDO parameters \label{app:trace_lpdo}}
The tensor elements in the LPDO $\Lambda_{\bm{\theta}}$ are initialized as random complex variables with gaussianly distributed real and imaginary part, namely
\begin{equation}
\label{eq:random_complex_vars}
\begin{split}
    & \theta_k = \Re(\theta_k) + i \Im(\theta_k) \\
    & \textrm{with}\,\, \Re(\theta_k),\, \Im(\theta_k) \sim \mathcal{N}(0, \sigma^2)\,.
\end{split}
\end{equation}

Given such choice, it is possible to compute the expectation value of the trace of the LPDO upon initialization, which amounts to 
\begin{equation}
\label{eq:exp_gaussian_trace_lpdo}
    \mathbb{E}_{\bm{\theta}}[\Lambda_{\bm{\theta}}] = (8\sigma^2\chi_\kappa)^n\,\chi_b^{n-1}\,,
\end{equation}
where $n$ is the number of qubits, and $\chi_\kappa$ and $\chi_b$ are the Kraus bond dimension and the virtual bond dimension of the LPDO, respectively. Then, by setting the variance to be $\sigma^2 = 2/(8\chi_\kappa\chi_b^{1 - 1 / n})$ one has that, in expectation value upon random initialization, the LPDO is properly normalized to the correct value $\mathbb{E}_{\bm{\theta}}[\Tr[\Lambda_{\bm{\theta}}]] = 2^n$. 

\begin{figure}[t]
    \centering
    \includegraphics[width=0.475\textwidth]{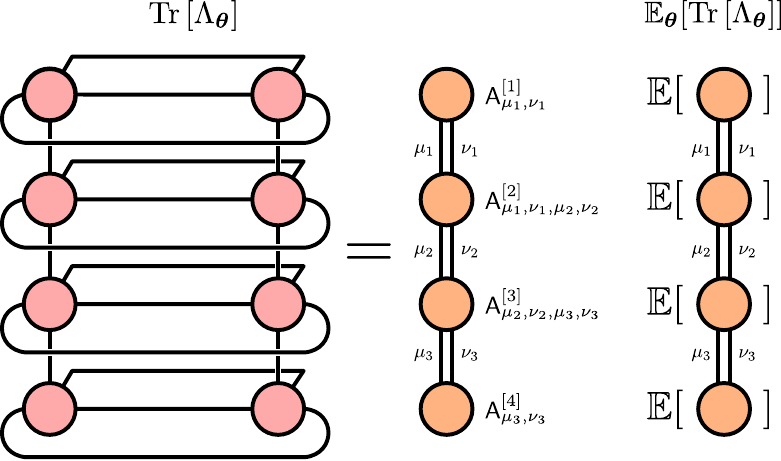}
    \caption{Left: tensor network representation of the trace of the LPDO parameterized by $\bm{\theta}$, which represents the noise channel. Right: tensor network representation of the expectation value of the trace of the LPDO upon initialization.}
    \label{fig:trace_lpdo}
\end{figure}

In what follows we show how to derive Eq.~\eqref{eq:exp_gaussian_trace_lpdo}, with the idea of the proof being diagrammatically depicted in Fig.~\ref{fig:trace_lpdo}. We first start by computing expectation values of the form $\mathbb{E}[\Tr[A\,A^\dagger]]$, where $A$ is a random matrix with normally distributed real and imaginary parts, and then proceed to show how the trace of the whole LPDO results in a composition of such quantities. 

Let $A \in \mathbb{C}^{2\times 2}$ be a complex random normal matrix whose entries are identically independently distributed (\textit{iid}) variables according to Eq.~\eqref{eq:random_complex_vars}. Then it holds
\begin{equation}
\label{eq:trace_random_mat}
    \begin{split}  
    \mathbb{E}\qty[\Tr[A\,A^\dagger]] & = 
    \mathbb{E}\qty[\Tr\qty[\begin{pmatrix} a   & b   \\ c   & d \end{pmatrix}
                           \begin{pmatrix} a^* & c^* \\ b^* & d^* \end{pmatrix}]] \\
    & = \mathbb{E}\qty[\abs{a}^2 + \abs{b}^2 + \abs{c}^2 + \abs{d}^2] \\
    & = 4\,\mathbb{E}\qty[\abs{a}^2] = 4\,\mathbb{E}\qty[\Re(a)^2 + \Im(a)^2] \\
    & =  8\sigma^2\,,
    \end{split}
\end{equation}
where in the third line we first made use of the fact that the matrix elements are \textit{iid}, and secondly that the real and imaginary parts satisfy $\mathbb{E}\qty[\Re(a)^2] = \mathbb{E}\qty[\Im(a)^2] = \sigma^2$. If instead one considers the product of two different independent random matrices $A$ and $B$, then it is easy to show that $\mathbb{E}[\Tr[AB]] = 0$.

We now proceed computing the trace of the LPDO tensor network $\Tr[\Lambda_{\bm{\theta}}]$, which is diagrammatically shown in Fig.~\ref{fig:trace_lpdo}(a). Let $A^{[q]}_{\mu, \nu, \kappa} \in \mathbb{C}^{2\times 2}$ denote the local Kraus tensors acting on each site, then starting from the definition of the LPDO given in Eq.~\eqref{eqn:LPDOaction} and tracing over the physical indices 
at each site of the tensor network results in
\begin{align}
\label{eqn:trace_lpdo}
     \phantom{=} &\Tr\qty[\Lambda_{\bm{\theta}}] = \nonumber\\[0.5em]
     = &\sum_{\bm{\mu},\bm{\nu},\bm{\kappa}} \textrm{Tr}\bigg[\bigg(A_{\mu_1,\kappa_1}^{[1]}\otimes A_{\mu_{1},\mu_2,\kappa_2}^{[2]} \otimes \ldots \otimes A_{\mu_{n-1},\kappa_n}^{[n]}\bigg) \times \nonumber\\
     & \quad\quad\quad \times \bigg(A^{[1]\dagger}_{\nu_1,\kappa_1}\otimes A_{\nu_{1},\nu_2,\kappa_2}^{[2]\dagger} \otimes \ldots\otimes  A_{\nu_{n-1},\kappa_n}^{[n]\dagger}\bigg)\bigg] \nonumber\\[0.5em]
     = & \sum_{\bm{\mu},\bm{\nu},\bm{\kappa}} \Tr[A_{\mu_1,\kappa_1}^{[1]}A^{[1]\dagger}_{\nu_1,\kappa_1}]  \Tr[A_{\mu_{1},\mu_2,\kappa_2}^{[2]}A_{\nu_{1},\nu_2,\kappa_2}^{[2]\dagger}] \times \nonumber\\ 
     & \quad\quad\quad\quad\quad\quad\quad\; \ldots \times \Tr[A_{\mu_{n-1},\kappa_n}^{[n]}A_{\nu_{n-1},\kappa_n}^{[n]\dagger}]\nonumber\\[0.5em]
     = & \sum_{\bm{\mu},\bm{\nu}} \mathsf{A}^{[1]}_{\mu_1,\nu_1} \mathsf{A}^{[2]}_{\mu_1,\nu_1, \mu_2, \nu_2} \ldots \mathsf{A}^{[n]}_{\mu_{n-1},\nu_{n-1}}\,,
\end{align}
where in the last line we introduced local tensors coming from the contraction of the Kraus indices $\bm{\kappa}$ and the trace over the physical indices at each site, namely
\begin{align}
\label{eq:trace_local_sites}
    \mathsf{A}^{[q]}_{\mu, \nu} &\coloneqq \sum_{\kappa=1}^{\chi_\kappa} \Tr\qty[A^{[q]}_{\mu, \kappa}\;A^{[q]\dagger}_{\nu, \kappa}]\quad\quad\,\, q \in \{1, n\}\,, \nonumber\\[-0.5em]
    \\[-0.5em]
    \mathsf{A}^{[q]}_{\mu, \nu, \mu', \nu'} &\coloneqq \sum_{\kappa=1}^{\chi_\kappa} \Tr[A_{\mu,\mu',\kappa}^{[q]}A_{\nu,\nu',\kappa}^{[q]\dagger}]\quad q \in [2, n-1]\,.\nonumber
\end{align}

When each matrix $A^{[q]}_{\mu, \nu, \kappa} \in \mathbb{C}^{2\times 2}$ in~\eqref{eq:single_site_trace} is a random matrix with normal complex entries, then one can use Eq.~\eqref{eq:trace_random_mat} to compute the expectation values
\begin{equation*}
    \mathbb{E}\qty[\mathsf{A}^{[q]}_{\mu, \nu}] = \sum_{\kappa=1}^{\chi_\kappa} \underbrace{\mathbb{E}\qty[\Tr\qty[A^{[q]}_{\mu, \kappa}\;A^{[q]\dagger}_{\nu, \kappa}]]}_{8\sigma^2\, \delta_{\mu\nu}} = 8\sigma^2\,\chi_\kappa\, \delta_{\mu\nu}\,,
\end{equation*}
and similarly for middle-sites tensors $q \in [2, n-2]$
\begin{equation*}
    \begin{split}
    \mathbb{E}\qty[\mathsf{A}^{[q]}_{\mu, \nu, \mu', \nu'}] &= \sum_{\kappa=1}^{\chi_\kappa} \underbrace{\mathbb{E}\qty[\Tr[A_{\mu,\mu',\kappa}^{[q]}A_{\nu,\nu',\kappa}^{[q]\dagger}]]}_{8\sigma^2\, \delta_{\mu\nu}\delta_{\mu'\nu'}}\\ 
    & = 8\sigma^2\,\chi_\kappa\, \delta_{\mu\nu}\delta_{\mu'\nu'}\,,
\end{split}
\end{equation*}
thus having in total
\begin{equation}
\label{eq:single_site_trace}
\begin{split}
    \mathbb{E}\qty[\mathsf{A}^{[q]}_{\mu, \nu}] &= 8\sigma^2\,\chi_\kappa\, \delta_{\mu\nu}\,,\phantom{\delta_{\mu'\nu'}} \quad q \in \{0, n-1\}\,, \\
     \mathbb{E}\qty[\mathsf{A}^{[q]}_{\mu, \nu, \mu', \nu'}] &= 8\sigma^2\,\chi_\kappa\, \delta_{\mu\nu}\delta_{\mu'\nu'}\,,\quad q \in [1, n-2]\,.
\end{split}
\end{equation}
Note that we can consider expectation values independently on each local tensor $\mathsf{A}^{[k]}$ since each tensor is drawn independently, and the expectation value factorizes over local sites, that is
\begin{align}
\label{eq:lpdo_full_trace}
    & = \mathbb{E}\qty[\Tr[\Lambda_{\bm{\theta}}]] \nonumber\\ 
    & = \mathbb{E}\qty[\sum_{\bm{\mu},\bm{\nu}} \mathsf{A}^{[1]}_{\mu_1,\nu_1} \mathsf{A}^{[2]}_{\mu_1,\nu_1, \mu_2, \nu_2} \ldots \mathsf{A}^{[n]}_{\mu_{n-1},\nu_{n-1}}] \\
    & = \sum_{\bm{\mu}, \bm{\nu}}\mathbb{E}\qty[\mathsf{A}^{[1]}_{\mu_{1}, \nu_{1}} ]\mathbb{E}\qty[\mathsf{A}^{[2]}_{\mu_{1}, \nu_{1}, \mu_{2}, \nu_{2}}]\hdots\, \mathbb{E}\qty[\mathsf{A}^{[n]}_{\mu_{n-1}, \nu_{n-1}}]\nonumber
\end{align}

Finally, using Eqs.~\eqref{eq:single_site_trace} inside Eq.~\eqref{eq:lpdo_full_trace}, one eventually obtains
\begin{align}
   & = \mathbb{E}\qty[\Tr[\Lambda_{\bm{\theta}}]]\nonumber\\ 
   & = \sum_{\bm{\mu}, \bm{\nu}}\,(8\sigma^2\chi_\kappa)^n\, \delta_{\mu_{1}\nu_{1}}\delta_{\mu_{1}\nu_{1}}\delta_{\mu_{2}\nu_{2}} \,\ldots\, \delta_{\mu_{n-1}, \nu_{n-1}}\nonumber\\
   & = (8\sigma^2\chi_\kappa)^n \chi_b^{n-1}\,,
\end{align}
where in the last line we used
\begin{align}
     & \sum_{\substack{\mu_1, \mu_2, \ldots, \mu_{n-1} = 1 \\ \nu_1, \nu_2, \ldots, \nu_{n-1} = 1}}^{\chi_b} \delta_{\mu_{1}\nu_{1}}\delta_{\mu_{1}\nu_{1}}\delta_{\mu_{2}\nu_{2}} \,\ldots\, \delta_{\mu_{n-1}, \nu_{n-1}}\nonumber\\
    & = \sum_{\substack{\mu_1, \mu_2, \ldots, \mu_{n-1} = 1 \\ \phantom{\nu_1,} \nu_2, \ldots, \nu_{n-1} = 1}}^{\chi_b}  \delta_{\mu_{2}\nu_{2}} \,\ldots\, \delta_{\mu_{n-1}, \nu_{n-1}}\\
    & = \sum_{\mu_1, \ldots, \mu_{n-1} = 1}^{\chi_b}  \phantom{1} = \chi_b^{n-1}\, .\nonumber
\end{align}

\subsection{Optimization details \label{app:training_details}}
The parameterized LPDO $\Lambda_{\bm{\theta}}$~\eqref{eqn:LPDO_definition} is trained with Adam optimizer~\cite{Kingma2017adam} combined with an exponential decay of the learning rate, which was found to stabilize training and ensure good convergence towards the end of the optimization process. 

Adam is a variant of stochastic gradient descent very common in machine learning research, and consists of the following update rules
\begin{equation}
    \begin{split}
        m_t & \leftarrow \beta_1  m_{t-1} + (1-\beta_1) g_t \\
        v_t & \leftarrow \beta_2  v_{t-1} + (1-\beta_2)  g_t ^ 2\\
        \hat{m}_t & \leftarrow m_t/(1-\beta_1^t) \\
        \hat{v}_t & \leftarrow v_t/(1-\beta_2^t) \\
        \theta_t & \leftarrow \theta_{t-1} - \eta  \hat{m}_t / \qty(\sqrt{\hat{v}_t} + \varepsilon)\,,
    \end{split}
\end{equation}
where $g_t = \nabla_{\theta}f(\theta_{t-1})$ is the gradient of the loss function $f(\theta)$ to be minimized having tunable parameters $\theta$, $g^2_t$ indicates its element-wise square, and $\eta$ is the step size (or learning rate). In our simulations we used standard values for the hyperparameters, $\beta_1 = 0.9$ and $\beta_2 = 0.999$, $\varepsilon=10^{-8}$. 

In addition to Adam, we used an exponential decay of the learning rate 
\begin{equation}
    \eta_t = \eta_0  \gamma ^ {t/T}
\end{equation}
where $\eta_0$ is the initial learning rate at the start of training, $\gamma$ is the decay rate, $t$ is the time step, and $T$ is a decay time. In our simulations we used $\eta_0 = 10^{-2}$, $\gamma = 0.9$, and the decay time $T$ was set equal to the number of training batches in an epoch, which depends on the number of tomographic samples $N$. The exponential decay stars only after a warm-up period of 500 gradient-descent steps. 

Importantly, note that in our case the parameters to be optimized are elements of the Kraus operators~\eqref{eqn:LPDO_definition}, and they consist of complex variables. Accordingly, the cost function is minimized by taking steps in the direction of the conjugated gradient~\cite{JaxAutodiffCookbook}. All optimization runs, including Adam and the exponential decay of the learning rate, were implemented as provided by the \texttt{jax}-based optimization library \texttt{optax}~\cite{OptaxDeepmind2020jax}.

As customary in machine learning, training was run by gradient-descent updates on mini-batches of data of size 250 (50 when the number of tomographic samples is scarce $N=10^3$). Of the whole tomographic dataset consisting of $N$ measurement samples, $\min(N/10, 12500)$ of them were used as a test dataset to estimate the loss function. \mar{The stopping criterion used during training was to stop the optimization if the Frobenius distance between the optimized LPDO and the target one didn't change more than $10^{-7}$ over the last 5 training epochs. In the realistic case scenario where one does not have the target LPDO to compare with, one can instead monitor the loss function on the test set and stop training if this stops improving.}

\section{Inversion of the MPOs}
\label{app:mpo_inversion}
In order to run the tensor error mitigation technique discussed in Sec.~\ref{sec:errorMit}, it is necessary to be able to compute the inverse of MPOs representing the quantum channels \mar{(which, however, is not a valid quantum channel~\cite{Mangini2022})}. That is, given a matrix product operator $\Gamma$, one needs to find another operator $\Gamma^{-1}$ such that $\Gamma \Gamma^{-1} = \mathbb{I}$. As proposed in~\cite{Guo2023}, this can be done by minimizing the error
\begin{equation}
    \label{eq:inv_mpo_error}
    \Delta_{\bm{\phi}} = \norm{\Gamma~\Upsilon_{\bm{\phi}} - \mathbb{I}}^2_F\,
\end{equation}
where $\norm{\bm{\cdot}}_F^2$ is the squared Frobenius distance, $\Gamma$ the MPO to be inverted, and $\Upsilon_{\bm{\phi}}$ is a parameterized MPO the tensor elements $\bm{\phi}$ of which are tuned to approach $\Gamma^{-1}$. As shown in~\cite{Guo2023}, this minimization problem can be reduced to a quadratic problem in the local tensors, and then solved by sweeping over the sites and solving local systems of linear equations at each of them. 

In addition to such explicit method, the error term $\Delta_{\bm{\phi}}$ can also be minimized by variationally tuning the parameters by means of an optimizer. Indeed, in our simulations we noticed that a combined approach of these two methods provides better results, especially when the MPO to be inverted is not sparse and contains many non-zero but small entries, as is the case for the MPOs coming from the noise characterization procedure~\ref{sec:dataSampling}. Specifically, one can use a classical optimization routine to find
\begin{equation}
    \bm{\phi}_{\text{opt}} = \argmin{\Delta_{\bm{\phi}}} = \argmin_{\bm{\phi}}\norm{\Gamma~\Upsilon_{\bm{\phi}} - \mathbb{I}}^2_F\,,
\end{equation}
and the optimization task can be performed either \textit{globally} by minimizing all the parameters in $\Upsilon_{\bm{\phi}}$ at the same time, or again in a DMRG-like~\cite{SchollwckDMRG2011} fashion by dividing it into many local subsequent optimization problems where only the parameters belonging to one single site are optimized at each time, with the others being fixed. 

For the tensor error mitigation experiments on $n=10$ qubits with the sparse Pauli-Lindblad noise reported in Sec.~\ref{sec:errorMit}, the exact noise maps ---that is those built explicitly from the definition of the noise channels--- were inverted with an MPO with virtual bond dimension $\chi_{b}=4$ using the linear algebra based inversion procedure proposed in~\cite{Guo2023}, which was found to converge to negligible inversion error $\Delta_{\bm{\phi}} \lessapprox 10^{-5}$. Instead, for the MPOs associated with the noise channels coming from the characterization procedure, the explicit inversion method, again with an ansatz MPO of bond dimension $\chi_{b}=4$, converged to $\Delta_{\bm{\phi}} \approx 6$, and was followed by a round of global and local variational minimization of the error function with optimizer L-BFGS-B as provided by \texttt{quimb}~\cite{GrayQuimb2018}, which improves the inversion achieving a final error of $\Delta_{\bm{\phi}} \approx 0.6$. 

Even though the inversion of the MPOs is not perfect, especially for the characterized noise channels, we note that in our cases the error from the inversion procedure is usually much smaller than the one from the characterization procedure, as one can see by comparing the normalized inversion error $\Delta_{\bm{\phi}_{\text{opt}}} / 2^{2n} \approx 10^{-7}$ with the normalized characterization error $\Delta(\Lambda, \Lambda_{\bm{\theta}_{opt}}) \approx 10^{-4}$ (see Fig.~\ref{fig:acc_vs_shots} with $10^7$ shots). We leave a more comprehensive analysis of the inversion error and their impact of noise mitigation as a topic for future studies. 

\bibliography{biblio}

\end{document}